\documentclass{lmcs}

\usepackage{import}
\usepackage{latexsym}
\usepackage{amssymb} 
\usepackage{amsmath} 
\usepackage{stmaryrd}
\usepackage[bbgreekl]{mathbbol}
\usepackage{url}
\usepackage{xypic}
\usepackage{tikz-cd}
\usepackage[T1]{fontenc}
\usepackage{breakurl} 
\usepackage{ebproof} 
\usepackage{scalerel,stackengine} 
\usepackage{hyperref}

\newcommand{\blank}{\ensuremath{{\mbox{-}}}}
\newcommand{\doubledownarrow}{\rotatebox[origin=c]{-90}{$\twoheadrightarrow$}}

\newcommand{\pfin}{\ensuremath{\powerset_{\mathrm{fin}}}}
\newcommand{\sembrack}[1]{\llbracket{#1}\rrbracket}
\newcommand{\upset}{\mathop{\uparrow}}
\newcommand{\dom}{\ensuremath{\mathrm{dom}}}
\newcommand{\uu}{\rotatebox[origin=c]{90}{$\twoheadrightarrow$}}
\newcommand{\Set}{\cat{Set}}
\newcommand{\Mble}{\ensuremath{\mathcal{M}ble}}

\newcommand{\Id}{\ensuremath{\mathrm{Id}}}
\newcommand{\Ell}{\mathcal{L}}

\newcommand{\cat}[1]{\ensuremath{\mathbf{#1}}}

\newcommand{\Ba}{\mathcal{B}a} 
\newcommand{\op}{\ensuremath{^{\mathrm{op}}}}
\newcommand{\Var}{\ensuremath{\mathrm{Var}}}
\newcommand{\R}{\ensuremath{\mathbb{R}}}
\newcommand{\Ord}{\cat{Ord}}
\newcommand{\dd}{\doubledownarrow}
\newcommand{\PMble}{\ensuremath{\mathcal{PM}ble}} 
\newcommand{\id}{\ensuremath{\mathrm{id}}}
\newcommand{\parfun}{\rightharpoonup}

\newcommand{\sigmaBA}{\cat{\sigma\text{-}BA}}
\newcommand{\N}{\ensuremath{\mathbb{N}}}
\newcommand{\im}{\mathrm{im}} 

\newcommand{\En}{\mathcal{N}}
\newcommand{\powerset}{\ensuremath{\mathcal{P}}}
\newcommand{\Bo}{\ensuremath{\mathcal{B}o}} 

\newcommand{\Zero}{\mathcal{Z}}

\newcommand{\princip}[1]{\left\langle #1 \right\rangle}
\newcommand{\ban}[1]{\llparenthesis #1 \rrparenthesis}
\newcommand{\Sik}{\mathcal{S}}
\newcommand{\sigmaBASp}{\sigmaBA_{\mathrm{sp}}}

\newcommand{\If}{\mathbf{if}}
\renewcommand{\Zero}{\mathbf{0}^?}
\newcommand{\Num}[1]{\mathbf{#1}^?}
\renewcommand{\succ}{\mathbf{succ}}
\newcommand{\pred}{\mathbf{pred}}

\newcommand{\lamth}{\boldsymbol{\lambda}} 

\newcommand{\Tot}[1]{\mathrm{Tot}} 
\newcommand{\Lang}{\mathfrak{L}}
\newcommand{\apowerset}{\powerset^A}
\newcommand{\axpowerset}{\powerset^{A(X)}}
\newcommand{\apfin}{\apowerset_{\mathrm{fin}}}
\newcommand{\axpfin}{\axpowerset_{\mathrm{fin}}}
\newcommand{\atimes}{\mathrel{\times_{\!A}}}

\stackMath
\newcommand{\verywidecheck}[1]{
	\savestack{\tmpbox}{\stretchto{
			\scaleto{
				\scalerel*[\widthof{\ensuremath{#1}}]{\kern-.6pt\bigvee\kern-.6pt}
				{\rule[-\textheight/2]{1ex}{\textheight}}
			}{\textheight}
		}{0.5ex}}
	\stackon[1pt]{#1}{\tmpbox}
}
\newcommand{\homoid}{\multimap} 
\newcommand{\ahom}[2]{\{ #1 \overset{A}{\rightarrow} #2 \}}

\newcommand{\beq}{\|\mbox{$=$}\|}
\newcommand{\bleq}{\|\mbox{$\leq$}\|}

\newcommand{\inn}{\varepsilon}
\newcommand{\graph}{\pmb{\gamma}}
\newcommand{\ungraph}{\pmb{\mathcal{F}}}

\newcommand{\SetoidR}{\cat{SetoidR}}
\newcommand{\SetoidF}{\cat{SetoidF}}

\newcommand{\Oid}{\mathrm{Oid}}

\newcommand{\scottcon}[2]{{[#1 \rightarrow #2]}}

\newcommand{\fv}{\mathrm{fv}}
\newcommand{\fun}{\mathbf{fun}}
\newcommand{\lam}{\mathbf{lam}}

\newcommand{\Const}{\mathfrak{K}}

\newcommand{\form}[1]{\mathrm{#1}} 
\newcommand{\hInductive}[1]{{#1}\text{-}\form{inductive}}

\newcommand{\iver}[1]{{\pmb{[}#1\pmb{]}}} 

\newcommand{\RI}{\mathcal{RI}} 

\newcommand{\bnot}{\lnot} 

\newcommand{\defeq}{\mathrel{\triangleq}}

\newcommand{\isoarrow}{\stackrel{\sim}{\rightarrow}}

\newcommand{\bvs}[1]{\| {#1} \|}

\newcommand{\cerp}{\mathbin{\reflectbox{$\prec$}}}

\newcommand{\symdiff}{\mathbin{\triangle}}

\begin{document}

\title[Interpreting $\lambda$-Calculus in Random Variables]{Interpreting Lambda Calculus in Domain-Valued Random Variables} 

\author[R. Furber]{Robert Furber\lmcsorcid{0000-0001-6208-9782}}[a]
\author[R. Mardare]{Radu Mardare\lmcsorcid{0000-0001-8660-1832}}[a]
\author[P. Panangaden]{Prakash Panangaden}[b]
\author[D. Scott]{Dana Scott}[c]

\address{Heriot-Watt University, Scotland}
\email{r.furber@hw.ac.uk, r.mardare@hw.ac.uk}

\address{McGill University, Canada}
\email{prakash@cs.mcgill.ca}

\address{Carnegie Mellon University, USA}
\email{scott@andrew.cmu.edu}

\begin{abstract}
  We develop Boolean-valued domain theory and show how the lambda-calculus can be
  interpreted using domain-valued random variables. We focus on the reflexive domain
  construction rather than the language and its semantics.  We develop the Boolean-valued
  set theory needed from scratch and then develop Boolean-valued domain theory on top of
  that.  The notions of equality and partial order have to be given Boolean-valued
  interpretations; when we say that an equation is valid in the model we mean that its
  interpretation is the top element of the Boolean algebra.
\end{abstract}

\maketitle

\section{Introduction}
There has been burgeoning interest in probabilistic programming languages in the
last decade. 
The main motivation is building \emph{compositional} models of probabilistic
processes and performing inference on them \cite{Roy11,Goodman08,Wood14}. 
For machine learning applications the notion of
conditioning is fundamental and the striking results of
\cite{Ackerman11,Ackerman19} show that this is a subtle issue. 

The combination of probability and higher type programming has been both
technically challenging and important for the development of semantics for such
languages.  There are a variety of approaches based on probabilistic coherence
spaces~\cite{Ehrhard14} or cones~\cite{Ehrhard17}, quasi-Borel
spaces~\cite{Staton16,Heunen17,Vakar19} and Boolean-valued
models~\cite{Bacci18b} which was based on Dana Scott's vision~\cite{Scott14}.
Stochastic lambda-calculi have appeared~\cite{Borgstrom16,Dahlqvist20,Amorim21}
with important contributions to the understanding of probability theory at
higher type.  The work on quasi-Borel spaces gives a cartesian closed category
that can serve as the foundation for a typed higher-order probabilistic
programming language~\cite{Heunen17}.  

In \cite{Bacci18b} a Boolean-valued domain theory was
developed.  The idea is to use one of the standard set-theoretic models of the
$\lambda$-calculus but interpreted in a suitable Boolean-valued universe of
sets.  However, the basic domain theoretic definitions of directed set,
supremum, reflexive dcpo and continuity were interpreted as usual.  The theory
presented there is rather complicated and had artificial restrictions on the way
randomness was handled.  In the present paper a \emph{completely} Boolean-valued
point of view is adopted and even basic concepts, like equality and order, are
all interpreted in a Boolean-valued logic.  This leads to a much simpler theory
but still in line with the vision of \cite{Scott14}. The version of Boolean-valued domain theory used in the current paper supersedes \cite{Bacci18b} by working ``internally'', by transferring theorems from ordinary logic to Boolean-valued logic as much as possible, rather than the ``bare handed'' approach of \cite{Bacci18b}. 

We show that untyped $\lambda$-calculus can be interpreted in domain-valued
random variables. In order to employ the theory of Boolean-valued sets, for most
of the article we actually use the Boolean-valued power set rather than random
variables, and then show that there is an isomorphism between the two at the
end.  We focus on the reflexive domain construction rather than the language and
its semantics. 

We are then able to generalize this way of doing things to random variables valued in an arbitrary dcpo with countable base.

The main contribution of this paper is the completely Boolean-valued reconstruction of domain
theory.  The
notion of equality has to be interpreted in the Boolean algebra and when we say
that an equation is valid in the model we mean that its interpretation is the
top element of the domain.

What makes the theory of Boolean-valued sets necessary is that domain-valued
random variables do not form a \emph{continuous} dcpo. However, when
``continuous dcpo'' is given its interpretation in Boolean-valued sets, they are.  
The version of Boolean-valued sets we have used is the original version in terms
of a cumulative hierarchy. If the reader prefers, they may rephrase the
arguments results in terms of topos theory in a Boolean topos, and we outline
the connection of the two in Section~\ref{SetoidSection}, though the only part
of this theory that we use is the ability to define functions.  
As an example of how to apply Boolean-valued domain theory, we show that there exist two sets of integers, neither of which can be mapped to the other by a $\lambda$-definable function. The reason for choosing this example is that it shows that $\lambda$-calculus and its model based on domain-valued random variables are powerful enough to prove a fact that does not mention probability in its statement, and such that the proof is pure $\lambda$-calculus and probability and doesn't need to pass via the equivalence between $\lambda$-definability and general recursion and applying the Kleene-Post theorem \cite[2.2 Corollary 1]{kleene1954} 
that there are incomparable many-one degrees. The proof, however, is inspired by Spector's probabilistic proof \cite[Theorem 2]{spector1958}.

There are a number of directions for future work.  First, one can use this construction to give semantics to a $\lambda$-calculus extended with probabilistic choice as was done in \cite{Bacci18b}.  In this model it will be interesting to see which equations involving the interplay of choice and the standard $\lambda$-calculus constructions are valid.  One could then relate it to an operational semantics as, for example, in \cite{Amorim21} but one would need a Boolean-valued notion of operational semantics.  More interestingly one could define conditioning as a primitive and explore its semantics.

 A second line of research is the notion of approximation.  A notion of ``approximate equality'' has been developed recently \cite{Mardare16}; the connection to the notion of equality used in the present paper is unclear but there is a similarity in that in both cases equality may only hold partially.

\section{Background on Boolean-Valued Set Theory}
\label{FirstOrderSection}
In this section we describe how statements and proofs in ordinary set theory can
be re-interpreted in a Boolean-valued sense. This interpretation is originally
due to Scott and Solovay~\cite{Scott67,jech,bell}. 

Throughout this section we let $A$ be an arbitrary complete Boolean algebra. It will play the role that the two-element Boolean algebra $2$ plays in ordinary logic. We can build up the class of $A$-valued sets, $V^A$, by considering an $A$-valued set $X$ to be a partial function that assigns to each element $x$ of its domain the amount, valued in $A$, that $x$ is an element of $X$. We build up $V^A$ by the following generalization of von Neumann's construction:
$$	V^A_0 = \emptyset,~~~V^A_{\alpha+1} = \{ f : V^A_\alpha \parfun A \},~~~
	V^A_{\alpha} = \bigcup_{\beta < \alpha}V^A_\beta ~~ \text{for } \alpha \text{ a limit ordinal.}
$$
Then either informally, or using proper classes, $V^A = \bigcup\limits_{\alpha \in \Ord} V^A_\alpha$.

We use the term \emph{$A$-valued set} for the elements of $V^A$, but we also will use the shorter term \emph{$A$-set}, and this helps to avoid certain confusions arising from the term ``valued''. 

We can then interpret $\in$, $\subseteq$ and $=$ by the following mutually recursive formulas: 
\begin{align*}
	\|x \in X\| = \bigvee_{t \in \dom X} \|x = t\| \land X(t) &&
	\|X \subseteq Y \| = \bigwedge_{t \in \dom X} X(t) \Rightarrow \|t \in Y\| \\
	\|x = y\| = \|x \subseteq y\| \land \|y \subseteq x\|.
\end{align*}
In the above, and all that follows, we use an operator precedence convention for $\bigvee$ and $\bigwedge$ that agrees with operator precedence for $\exists$ and $\forall$, so there is an implicit bracket around everything to the right of such a join or meet.

Recall that the first-order language of set theory, which we write as $\Lang_\Set$, is the usual first-order language for a signature with equality and one two-place relation symbol, namely $\in$. We write $\Lang_\Set(V^A)$ for this language extended with constants from $V^A$. 

In any Boolean algebra, we can interpret the connectives of propositional logic. Using the completeness of $A$, we can interpret the universal quantifier as a meet and the existential quantifier as a join. All together, this gives us an interpretation of $\Lang_\Set(V^A)$ in $V^A$. 

\begin{thm}[$V^A$ as a model]
	\label{SetTheoryMetaTheorem}\hfill
	\begin{enumerate}[(i)]
		\item If $\phi \in \Lang_\Set(V^A)$ is a theorem of ZFC set theory, then $\| \phi \| = 1$ in $V^A$.
		\item The inference rules of first-order logic can be applied to theorems of ZFC set theory and statements about elements of $V^A$ in $\Lang_\Set(V^A)$. 
		\item If $\| \exists x. \Phi(x, X_1, \cdots, X_n) \|  = a$, where $X_1, \ldots, X_n \in V^A$ (and we allow $n = 0$), then there exists $X_0 \in V^A$ such that $\| \Phi(X_0, X_1, \cdots, X_n) \| = a$. 
	\end{enumerate}
\end{thm}
\begin{proof}
For (i) and (ii), see \cite[Theorem 1.17]{bell}; (iii) is a particular instance of what we shall later call the ``completeness'' of $V^A$ - see \cite[Lemma 14.19]{jech} or \cite[Lemma 1.27]{bell}. 
\end{proof}

It is helpful to make certain constructions of $A$-sets explicit. If $x,y \in V^A$, we can define the singleton $\{x\}^A$, unordered pair $\{ x,y \}^A$ and ordered pair $(x,y)^A$ as follows:
\begin{align*}
	\dom\{x\}^A &= \{ x \}  & \dom(\{x,y\}^A) &= \{ x, y\} \\
	\{x\}^A(x) &= 1 & \{x,y\}^A(x) &= 1 \\
	&     & \{x,y\}^A(y) &= 1 \\
	\dom((x,y)^A) &= \{ \{x\}^A, \{x,y\}^A \} \\
	(x,y)^A(\{x\}^A) &= 1 \\
	(x,y)^A(\{x,y\}^A) &= 1.
\end{align*}
For $X,Y \in V^A$ we define $X \atimes Y$ as follows.

$X \atimes Y : \{ (x,y)^A \mid x \in \dom(X), y \in \dom(Y) \} \rightarrow A$

	$(X \atimes Y)(x,y)^A = \| x \in X \| \land \| y \in Y \|.$
\\For each $A$-set $X$, the $A$-valued power set $\apowerset(X)$ is defined by

	$\dom(\apowerset(X)) = \{ u \in V^A \mid \dom(u) = \dom(X) \text{ and } \forall t \in \dom(u) . u(t) \leq X(t) \}$ 
	
	$\apowerset(X)(u) = 1$
\\For all $S, X \in V^A$ we have
$
\| S \in \apowerset(X) \| = \| S \subseteq X \|,
$
which is how the power set axiom is proved for $V^A$. 
\\If $\Phi$ is a set-theoretic formula we can define the $A$-set
$$\{ x \in X \mid \Phi(x) \}^A : \dom(X) \rightarrow A~~\text{ by}$$ $$\{ x \in X \mid \Phi(x) \}^A(t) = X(t) \land \| \Phi(t) \|,$$
which proves the axiom of separation for $V^A$. 

If we now consider von Neumann's universe $V$ of classic set theory constructed inductively on ordinals $\alpha \in \Ord$, then for each set $S \in V_\alpha$, there is a corresponding element $\check{S} \in V^A_\alpha$ defined recursively. The domain of $\check{S}$ is $\{\check{x} \mid x \in S \}$, and it is defined by:
$$	\check{\emptyset} = \emptyset,~~~\text{and }~
	\check{S}(\check{x}) = 1 ~ \text{for all } x \in S.$$
To describe how a statement in set theory about an ordinary set in $X \in V$ translates to a statement about $\check{X} \in V^A$, we need the notion of a $\Delta_0$ statement. Bounded quantifiers are those of the form $\forall x \in X. \Phi(x)$ and $\exists x \in X . \Psi(x)$, \emph{i.e.} in the language of set theory $\forall x. x \in X \Rightarrow \Phi(x)$ and $\exists x . x \in X \land \Psi(x)$. A $\Delta_0$ formula $\Phi \in \Lang_\Set(V^A)$ is one containing only bounded quantifiers. We have $\Delta_0$ invariance\footnote{Note that Bell uses the alternative terminology ``restricted'' for $\Delta_0$.} \cite[Lemma 14.21]{jech} or \cite[Theorem 1.23 (v)]{bell}.

\begin{thm}[$\Delta_0$ invariance]
	\label{DeltaNoughtMetaTheorem}
	Let $\Phi(x_1,\ldots,x_n)$ be a $\Delta_0$-formula, whose free variables are $x_1, \ldots, x_n$. Let $X_1,\cdots,X_n \in V$. Then,
	$
	\Phi(X_1,\ldots,X_n) \Leftrightarrow \| \Phi(\check{X}_1, \ldots, \check{X}_n) \| = 1.
	$
\end{thm}

We point out some useful consequences of Theorem \ref{DeltaNoughtMetaTheorem}. The statements ``$S$ is a singleton whose only element is $x$'', ``$S$ is an unordered pair of $x$ and $y$'' and ``$S$ is an ordered pair, first element $x$, second element $y$'' are all $\Delta_0$ statements, and so the following all hold
\begin{align*}
	\verywidecheck{\{x\}} &= \{\check{x}\}^A & \verywidecheck{\{x,y\}} &= \{ \check{x},\check{y} \}^A & \verywidecheck{(x,y)} &= (\check{x},\check{y})^A.
\end{align*}
Another important consequence of Theorem \ref{DeltaNoughtMetaTheorem} is that $\check{\omega}$ is the smallest inductive\footnote{In the sense used to formulate the axiom of infinity in ZF.} $A$-set in $V^A$, so that $\check{\omega}$ is the $\omega$ of $V^A$. Given any set $X$, we say a subset $S \subseteq X$ is finite if there exists $n \in \omega$ and $f : n \rightarrow X$ such that $S = \im(f)$. We write $\pfin(X)$ for the set of all finite subsets of $X$. If we interpret this with $A$-sets, for each $X \in V^A$ we can define $\apfin(X)$ to be the set of finite $A$-subsets of $X$, using the axiom of separation. The following proposition is a consequence of the distributive law - see \cite[3.1.11]{kusraev}.
\begin{prop}
	\label{FiniteSetPreservationProp}
	For all sets $X$, 
	$
	~~~\| \apfin(\check{X}) = \verywidecheck{\pfin(X)} \| = 1.
	$
\end{prop}

We can define the category $\Set_A$ from $V^A$ as follows. The class of objects is $V^A$, and for each $X,Y \in V^A$, the set of functions $\ahom{X}{Y} \in V^A$ can be defined using products, power set, and the axiom of separation. We define:
\begin{align*}
	\Set_A(X,Y) &= \{ f \in \dom(\ahom{X}{Y}) \mid \| f \in \ahom{X}{Y} \| = 1 \}/\sim,
\end{align*}
where $\sim$ is the equivalence relation defined by $f \sim g$ iff $\| f = g \| = 1$. Identity elements are defined as identity relations, and composition by composition of relations, which is well-defined with respect to $\sim$.


\section{Boolean-Valued Setoids}
\label{SetoidSection}
We now discuss the kind of ``Boolean-valued sets'' that are essentially $A$-valued models of the theory of $=$. These can be considered an alternative description of either sheaves or separated presheaves on $A$, and are usually discussed in the more general case where $A$ is a Heyting algebra, such as in \cite{fourman1979,goldblatt,monroA}. 
Although we will use only one definition of $A$-valued sets, the distinction between $A$-valued relations that satisfy the definition of a function \emph{internally} and functions that do so gives us two distinct categories. 

In order to distinguish the $A$-valued sets we will be discussing from elements of $V^A$, we call them \emph{$A$-valued setoids}\footnote{\emph{Setoid} is a long-established term for a set equipped with an equivalence relation.}, or simply \emph{$A$-setoids}.
An $A$-setoid is a pair $(X,\beq_X)$ where $X$ is a set, $\beq_X : X \times X \rightarrow A$. The two axioms of $\beq_X$ are symmetry and transitivity, \emph{i.e.}
\begin{align*}
	\forall x,y \in X . \| x = y \|_X &= \| y = x \|_X \\
	\forall x,y,z \in X. \| x = y \|_X \land \| y = z \|_X &\leq \| x = z \|_X.
\end{align*}
so $A$-valued setoids are ``$A$-valued \emph{partial} equivalence relations''. 

Since reflexivity is not assumed, the statement $\| x = x \|_X$ does not represent a tautology, but its intended interpretation is, in some sense, the degree to which $x \in X$, or the place where $x \in X$. So we define the notation $\inn_X(x) = \| x = x \|_X$. 

We say $(X, \beq_X)$ is \emph{total} if $\|x = x\|_X = 1$ for all $x \in X$, \emph{i.e.} if reflexivity holds; it is \emph{strict} if $\|x = y\|_X = 1$ implies $x = y$. 

\begin{defi}
	\label{CompletenessDef}
	Let $(X,\beq_X)$ be an $A$-setoid. We say it is \emph{complete}\footnote{This definition is slightly different from that used in \cite[4.10,4.11]{fourman1979} and elsewhere, which would not suit us because under that definition $\apowerset(X)$ is not complete, when considered as an $A$-setoid in the sense we will define in Definition \ref{SetsAreSetoidsDefinition}.} if either, hence both, of the following equivalent properties holds:
	\begin{enumerate}[(i)]
		\item For any family of elements $(a_i)_{i \in I}$ in $A$ and corresponding family $(x_i)_{i \in I}$ in $X$ such that for all $i,j \in I$, $a_i \land a_j \leq \|x_i = x_j\|_X$, there exists $x \in X$ such that for all $i \in I$, $a_i \leq \| x_i = x \|_X$. 
		\item For any pairwise disjoint family of elements $(a_i)_{i \in I}$ in $A$ and corresponding family $(x_i)_{i \in I}$ in $X$ such that for all $i \in I$, $a_i \leq \| x_i = x_i \|_X$, there exists $x \in X$ such that for all $i \in I$, $a_i \leq \| x_i = x\|_X$. 
	\end{enumerate}
\end{defi}

\begin{defi}
	\label{StrictIsoDef}
	If $(X,\beq_X)$ is an $A$-setoid and $(Y,\beq_Y)$ is a set and a function $\beq_Y : Y \times Y \rightarrow A$, and $f : X \rightarrow Y$ is a bijection such that for all $x_1, x_2 \in X$:
	$$
	\| f(x_1) = f(x_2) \|_Y = \| x_1 = x_2 \|_X,
	$$
	then we say $f$ is a \emph{strict isomorphism}, and it follows that $(Y,\beq_Y)$ is an $A$-setoid, and is total, strict or complete iff $(X,\beq_X)$ is.
\end{defi}

\begin{defi}
	\label{ProductSetoidDef}
	The product $(X,\beq_X) \times (Y,\beq_Y)$ is defined to have
	\[
	\| (x,y) = (x',y') \|_{X \times Y} = \| x = x' \|_X \land \| y = y' \|_Y.
	\]
\qedhere
\end{defi}

\begin{defi}
	\label{PredicateDef}
	A \emph{predicate} on $(X,\beq_X)$ is a function $S : X \rightarrow A$ such that:
	\begin{enumerate}[(i)]
		\item For all $x_1, x_2 \in X$, $\| x_1 = x_2 \|_X \leq S(x_1) \Leftrightarrow S(x_2)$. 
		\item For all $x \in X$, $S(x) \leq \inn_X(x)$.
	\end{enumerate}
A binary relation $X \rightarrow Y$ is simply a predicate on $X \times Y$. \qedhere
\end{defi}
This allows us to define a category of setoids with relations that satisfy the internal definition of a function (``relational functions'') as morphisms. 
\begin{defi}
	\label{SetoidRelDef}
	The category $\SetoidR_A$ has $A$-setoids as objects and a morphism is a relation $f: X \rightarrow Y$ such that (additionally to (i) and (ii) from Definition \ref{PredicateDef}):
 \begin{enumerate}[(i)]
		\setcounter{enumi}{2}
		\item For all $x \in X$, $y_1,y_2 \in Y$, $f(x,y_1) \land f(x,y_2) \leq \| y_1 = y_2 \|_Y$. 
		\item For all $x \in X$, $\inn_X(x) \leq \bigvee_{y \in Y} f(x,y)$. 
	\end{enumerate}
	The identity function $\id_X : X \rightarrow X$ is defined by:
	$
	\id_X(x_1,x_2) = \| x_1 = x_2\|_X.
	$
	\\ Composition of functions $f : X \rightarrow Y$ and $g : Y \rightarrow Z$ is defined by
	\[
	(g \circ f)(x,z) = \bigvee_{y \in Y} f(x,y) \land g(y,z). \qquad 
	\]
\qedhere
\end{defi}

We define the category of $A$-setoids and ``functional functions'', $\SetoidF_A$, as follows. 
\begin{defi}
	\label{SetoidFDef}
	The category $\SetoidF_A$ has $A$-setoids as objects. The morphisms are defined starting with an $A$-setoid:
	\begin{align*}
		X \homoid Y &= \{ f : X \rightarrow Y \mid \forall x_1,x_2 \in X. \|x_1= x_2\|_X \leq \|f(x_1) = f(x_2)\|_Y \} \\
		\| f_1 = f_2 \|_{X \homoid Y} &= \bigwedge_{x \in X} \inn_X(x) \Rightarrow \| f_1(x) = f_2(x) \|_Y.
	\end{align*}
	Then the hom set $\SetoidF_A(X,Y)$ is the quotient set of this,\footnote{Monro calls these \emph{admissible functions} in \cite[Definition 5.2]{monroA}.} \emph{i.e.} $X \homoid Y$ modulo the equivalence relation:
	\[
	f_1 \sim f_2 \text{ iff } \forall x \in X. \inn_X(x) \leq \| f_1(x) = f_2(x) \|_Y.
	\]
	Identity maps and composition are defined as for functions. \qedhere
\end{defi}

The category $\SetoidF_A$ is called $\cat{Mod}_0(A)$ in \cite[Definition 5.2]{monroA}, where it is proved to be a quasi-topos. The full subcategory on non-empty\footnote{We mean not having $\emptyset$ as an underlying set. It is still allowed, and necessary, to have sets that do not assign any value other than $0$ to any element, which are ``empty'' in terms of Boolean-valued logic.} complete $A$-setoids is a topos \cite[Proposition 5.6 (ii)]{monroA} equivalent to the category of sheaves on the canonical topology on $A$.

Mapping a functional function to its graph defines a faithful functor from $\SetoidF_A \rightarrow \SetoidR_A$.
\begin{defi}
	\label{FloorFunDef}
	The following defines a faithful functor $\graph : \SetoidF_A \rightarrow \SetoidR_A$. On objects, $\graph(X,\beq_X) = (X,\beq_X)$. For $f \in \SetoidF_A(X,Y)$, we define
	\[
	\graph(f)(x,y) = \inn_X(x) \land \| f(x) = y \|_Y.
	\]
\qedhere
\end{defi}

We can formulate a notion of poset that does not require us to define $\beq_X$ but can instead be defined in terms of it. It turns out to be useful in the general theory, as well as for dealing with posets that are $A$-setoids or $A$-sets. 

\begin{defi}
	\label{BVPosetDef}
	An \emph{$A$-poset} is a pair $(X,\bleq_X)$, where $X$ is a set and $\bleq_X : X \times X \rightarrow A$ is such that for all $x_1,x_2,x_3 \in X$:
	\begin{enumerate}[(i)]
		\item  $\| x_1 \leq x_2 \|_X \land \| x_2 \leq x_3 \|_X \leq \| x_1 \leq x_3 \|_X.$
		
		\item $ \| x_1 \leq x_2 \|_X \leq \| x_1 \leq x_1 \|_X \land \| x_2 \leq x_2 \|_X.$
	\end{enumerate}
	We then define $\beq_X$ by
	\begin{equation}
		\label{PosetEqDefEqn}
		\| x_1 = x_2 \|_X = \| x_1 \leq x_2 \|_X \land \| x_2 \leq x_1 \|_X.
	\end{equation}
	Then $(X,\beq_X)$ is an $A$-setoid, and $\bleq_X$ is a reflexive, antisymmetric, transitive relation when these statements are interpreted in the $A$-valued sense. 
	If $(X,\beq_X)$ is an $A$-setoid and $\bleq_X$ is a relation that is reflexive, antisymmetric and transitive, in the $A$-valued sense, then $(X,\bleq_X)$ is an $A$-poset in the above sense, and $\beq_X$ satisfies \eqref{PosetEqDefEqn}.
\end{defi}

Since $\beq_X$ is defined in terms of $\bleq_X$, the following fact is convenient for proving that a function on underlying sets $f : X \rightarrow Y$ not only defines an element of $\SetoidF_A(X,Y)$ but also a monotone function in the $A$-valued sense.

\begin{lem}
	\label{MonotoneImpliesFunctionLemma}
	Let $(X,\bleq_X), (Y,\bleq_Y)$ be $A$-posets, and suppose that $f : X \rightarrow Y$ is $A$-monotone in the sense\footnote{This is monotonicity using the interpretation of logic in Section \ref{FirstOrderSection}.} that for all $x_1,x_2 \in X$,
	$
	\| x_1 \leq x_2 \|_X \leq \| f(x_1) \leq f(x_2) \|_Y.
	$
	\\Then $f \in X \homoid Y$. Every $f \in \SetoidF_A(X,Y)$ that is monotone in the $A$-valued interpretation of the term is of this form.
\end{lem}

The following is useful for transferring the property of being an $A$-poset to a set equipped with an $A$-valued relation.
\begin{lem}
\label{BuildPosetByIsoLemma}
Let $(Y,\bleq_Y)$ be an $A$-valued poset, and $(X,\bleq_X)$ a pair such that $X$ is a set and $\bleq_X$ is a function $X \times X \rightarrow A$. Suppose that $f : X \rightarrow Y$ is a bijection such that for all $x_1,x_2 \in X$, $\bvs{x_1 \leq x_2}_X = \bvs{f(x_1) = f(x_2)}_Y$, then $(X,\bleq_X)$ is an $A$-poset, and the map $f$ and its inverse are $A$-monotone, and $f$ is a strict isomorphism of setoids.
\end{lem}
\begin{proof}
For all $x_1,x_2,x_3 \in X$,
\begin{align*}
\bvs{x_1 \leq x_2}_X \land \bvs{x_2 \land x_3}_X &= \bvs{f(x_1) \leq f(x_2)}_Y \land \bvs{f(x_2) \leq f(x_3)}_Y \\
 &\leq \bvs{f(x_1) \leq f(x_3)}_Y \\
 &= \bvs{x_1 \leq x_3}_X,
\end{align*}
and for all $x_1,x_2 \in X$
\begin{align*}
\bvs{x_1 \leq x_2}_X &= \bvs{f(x_1) \leq f(x_2)}_Y \\
 &\leq \bvs{f(x_1) \leq f(x_1)}_Y \land \bvs{f(x_2) \leq f(x_2)}_Y \\
 &= \bvs{x_1 \leq x_1}_X \land \bvs{x_2 \leq x_2}_X.
\end{align*}
So we have proved that $(X,\bleq_X)$ is an $A$-valued poset, and also that $f$ is $A$-valued monotone, and so is its inverse. We also have
\begin{align*}
\bvs{f(x_1) = f(x_2)}_Y &= \bvs{f(x_1) \leq f(x_2)}_Y \land \bvs{f(x_2) \leq f(x_1)}_Y \\
 &= \bvs{x_1 \leq x_2}_X \land \bvs{x_2 \leq x_1}_X \\
 &= \bvs{x_1 = x_2}_X,
\end{align*}
so $f$ is a strict isomorphism of $A$-setoids. 

To see that $f^{-1}$ is $A$-monotone, simply observe that for all $y_1,y_2 \in Y$
\[
\bvs{f^{-1}(y_1) \leq f^{-1}(y_2)}_X = \bvs{f(f^{-1}(y_1)) \leq f(f^{-1}(y_2))}_Y = \bvs{y_1 \leq y_2}_Y,
\]
and read it in reverse. 
\end{proof}

The following definition and theorem are based on \cite[Proposition 3.3]{monroB}. 
\begin{defi}
	\label{CeilFDef}
	Let $(X,\beq_X), (Y,\beq_X)$ be $A$-setoids such that $Y$ is complete, and let $f \in \SetoidR_A(X,Y)$. For each $x \in X$, there exists $\ungraph(f)(x) \in Y$ such that
	\begin{equation}
		\label{CeilDefiningPropertyEqn}
		f(x,y) \leq \| \ungraph(f)(x) = y \|_Y.
	\end{equation}
	Any choice of $\ungraph(f)(x)$ for all $x \in X$ defines an element $\ungraph(f) \in \SetoidF_A(X,Y)$ such that $\graph(\ungraph(f)) = f$, and in fact this establishes a $\SetoidF_A$ isomorphism $\SetoidR_A(X,Y) \cong \SetoidF_A(X,Y)$. \qedhere
\end{defi}

Every $A$-set defines an $A$-setoid as follows.

\begin{defi}
	\label{SetsAreSetoidsDefinition}
	Let $X \in V^A$. Define $\beq_X : \dom(X)\times\dom(X) \rightarrow A$ by
	\begin{equation}
		\label{SetSetoidEqn}
		\| x = y \|_X = \| x \in X \| \land \|y \in X \| \land \|x = y\|
	\end{equation}
	Then $(\dom(X), \beq_X)$ is an $A$-setoid, which we write $\Oid(X)$ when we need to be unambiguous. When we speak of elements of $\Oid(X)$, we mean elements of the underlying set of $\Oid(X)$, \emph{i.e.} elements of $\dom(X)$, following the usual convention for sets with structure. \qedhere
\end{defi}

\begin{defi}
	\label{OidSubsetDef}
	Let $X \in V^A$. For each $S \in V^A$ such that $\| S \subseteq X \| = 1$, then $e(S)$, defined as follows, is a predicate on $\Oid(X)$:
	\[
	e(S)(x) = \| x \in S \|,
	\]
	where $x \in \Oid(X)$. In particular, this holds if $S \in \dom(\apowerset(X))$. \qedhere
\end{defi}

\begin{defi}
	\label{OidFunctorDef}
	The following defines $\Oid : \Set_A \rightarrow \SetoidR_A$ a functors, where $f : X \rightarrow Y$ is in $\Set_A$:
	\[
	\Oid(f)(x,y) = \| (x,y)^A \in f \|
	\]
	This functors is full, faithful and essentially surjective, so $\Set_A \simeq \SetoidR_A$. \qedhere
\end{defi}

\begin{thm}
	\label{PowerSetCompletenessTheorem}
	For each $X \in V^A$, $\Oid(\apowerset(X))$ is complete. Moreover, if $\Phi$ is a formula of $\Lang_\Set(V^A)$ and $\| \exists S \in \apowerset(X) . \Phi(S,\ldots) \| = a \in A$, then there exists $S \in \Oid(\apowerset(X))$ such that $\| \Phi(S,\ldots)\| = a$. \qedhere
\end{thm}

\begin{cor}
	\label{RelFuncExistenceCor}
	For each $X,Y \in V^A$, $\Oid(\apowerset(X \atimes Y))$ and $\Oid(\ahom{X}{Y})$ are complete, and if we prove a relation or a function exists in the Boolean-valued sense, satisfying some formula $\Phi \in \Lang_\Set(V^A)$, then there exists a corresponding element of $\Oid(\apowerset(X \atimes Y))$ or $\Oid(\ahom{X}{Y})$ satisfying $\Phi$. \qedhere
\end{cor}


\section{Models of Untyped Lambda-calculus in Boolean-valued sets}
\label{ClassicalLambdaSection}
In this section we show how to model untyped $\lambda$-calculus in $A$-valued sets. We use the Engeler model \cite{engeler1981} as the main example. We will repeatedly use Theorem \ref{SetTheoryMetaTheorem} to prove statements in $V^A$ by proving them in set theory first. The reader who would prefer to use a topos formulation might prefer to use the categorical formulation of the Engeler model from \cite{hyland2006}. 

\subsection{Background on Domain Theory}
A \emph{dcpo} $(D,\leq)$ is a poset that is directed complete, \emph{i.e.} such that every directed set has a supremum. Every complete lattice is a dcpo. The least element of a dcpo $(D,\leq)$, if it exists, is called the \emph{bottom element} and is written $\bot$. 
In a dcpo $D$, where $d,e \in D$, we say that $d$ is \emph{way below} $e$, written $d \ll e$, if for each directed set $(e_i)_{i \in I}$ such that $e \leq \bigvee_{i \in I}e_i$, there exists some $j \in I$ such that $d \leq e_j$. The relation $\ll$ is transitive and antisymmetric, but in general is neither reflexive nor irreflexive.
We write $\dd d = \{ e \in D \mid e \ll d \}$. 
A dcpo $D$ is called \emph{continuous} if for all $d \in D$, the set $\dd d$ is directed and $d = \bigvee \dd d$. 
A continuous dcpo is called a \emph{domain}. A complete lattice that is continuous as a dcpo is called a \emph{continuous lattice}.
A \emph{basis} or \emph{base} for a domain $D$ is a subset $B \subseteq D$ such that for all $d \in D$, $\dd d \cap B$ is directed, and $d = \bigvee ( \dd d \cap B)$. It follows that if $D$ is continuous, then $D$ is a base for $D$. If $D$ has a countable base, we say it is a \emph{countably-based domain}. See \cite{gierz} for more related concepts.

If $D$ and $E$ are dcpos, a function $f : D \rightarrow E$ is said to be \emph{Scott continuous} if it is monotone and preserves directed suprema. We write $\scottcon{D}{E}$ for the set of Scott-continuous maps $D \rightarrow E$. This is a dcpo when given the pointwise ordering, and directed suprema are calculated pointwise. When we refer to this as a dcpo, we always use this structure. 
The \emph{Scott topology} on a dcpo $D$ is the topology whose open sets are the sets $U \subseteq D$ such that $U$ is an up set and for all directed sets $(d_i)_{i \in I}$ such that $\bigvee_{i \in I}d_i \in U$ there exists $i \in I$ such that $d_i \in U$. If $D$ and $E$ are dcpos, a function $f : D \rightarrow E$ is Scott continuous iff it is a continuous map from $D$ to $E$ equipped with their respective Scott topologies. 


\begin{defi}
	\label{ReflexiveDcpoDef}
	A \emph{reflexive dcpo} 
	is a triple $(D,\fun,\lam)$, where $D$ is a dcpo with bottom\footnote{This is what is meant by a cpo in \cite[Definition 1.2.1 (ii)]{barendregt84}.}, and $\fun : D \rightarrow \scottcon{D}{D}$ and $\lam : \scottcon{D}{D} \rightarrow D$ are Scott-continuous maps making $\scottcon{D}{D}$ a retract of $D$, \emph{i.e.} $\fun \circ \lam = \id_{\scottcon{D}{D}}$. A reflexive dcpo is \emph{extensional} iff $\fun$, or equivalently $\lam$ is an isomorphism, \emph{i.e.} iff additionally $\lam \circ \fun = \id_D$. \qedhere
\end{defi}
It is often convenient to formulate this notion using the uncurried form of $\fun$, which is to say, we can equivalently define a reflexive dcpo to be a triple $(D,\cdot, \lam)$ where $D$ is a dcpo with bottom, and $\cdot : D \times D \rightarrow D$ and $\lam : \scottcon{D}{D} \rightarrow D$ are Scott-continuous maps such that for each $f \in \scottcon{D}{D}$ and $d \in D$ we have $\lam(f) \cdot d = f(d)$. 

The language of untyped $\lambda$-calculus can be interpreted in a reflexive dcpo as follows. Given a reflexive dcpo $(D,\cdot,\lam)$, and a set of variables $\Var$, a \emph{valuation} 
is a partial function $\rho : \Var \rightarrow D$. We also choose a set of constants $\Const \subseteq D$, which is allowed to be any subset, including $\emptyset$ and $D$ itself. We then define the language of $\lambda$-calculus $\Lambda(D,\Var,\Const)$ as follows
\begin{defi}
	\label{LambdaCalcLangDef}
	The language $\Lambda(D,\Var,\Const)$ is defined inductively by the rules 
	\[
	\begin{prooftree}
		\hypo{x \in \Var}
		\infer1{x \in \Lambda(D,\Var,\Const)}
	\end{prooftree}\quad~~~~~~~~~~~~
	\begin{prooftree}
		\hypo{d \in \Const}
		\infer1{d \in \Lambda(D,\Var,\Const)}
	\end{prooftree}
	\]
	
	\[
	\begin{prooftree}
		\hypo{x \in \Var}\hypo{M \in \Lambda(D,\Var,\Const)}
		\infer2{\lambda x . M \in \Lambda(D,\Var,\Const)}
	\end{prooftree}\quad~~~~~~~~~~~~
	%
	\begin{prooftree}
		\hypo{M \in \Lambda(D,\Var,\Const)}\hypo{N \in \Lambda(D,\Var,\Const)}
		\infer2{MN \in \Lambda(D,\Var,\Const)}
	\end{prooftree}
	\]
\qedhere
\end{defi}
We write $\Lambda(\Var)$ for $\Lambda(D,\Var,\emptyset)$, since this does not depend on $D$. The elements of $\Lambda(\Var)$ are called \emph{pure $\lambda$-terms}. We write $\fv(M)$ for the set of free variables of $M$, defined as usual.

We make the following observation about how the above definition is formulated in set theory\footnote{This is needed to make the proof of Proposition \ref{LambdaTermInternalProp} and later results that deal with viewing the syntax of $\lambda$-calculus from inside $V^A$ more comprehensible. }. 
\begin{exa}
	\label{SetTheoryFormulationExample}
	There exist $\Delta_0$ formulas $\hInductive{\Lambda(D,\Var,\Const)}$ and $\hInductive{\Lambda(\Var)}$, such that the sets $\Lambda(D,\Var,\Const)$ and $\Lambda(\Var)$ are respectively definable as the smallest sets satisfying their corresponding formula. 

Indeed, to define $\hInductive{\Lambda(\Var)}$ and $\hInductive{\Lambda(D,\Var,\Const)}$, we need a way to represent the syntax of $\lambda$-calculus. For example, we could take the ordinals $0$ to mean a variable, $1$ a $\lambda$-abstraction, $2$ an application and $3$ a constant, so that a variable appears as $(0,x)$ for $x \in \Var$, a $\lambda$-abstraction as $(1,(x,M))$ where $x \in \Var$ and $M$ is a $\lambda$-term, an application as $(2, (M,N))$ where $M,N$ are $\lambda$-terms, and a constant as $(4,d)$ where $d \in \Const$. 
	
Then we can define:
	\begin{align*}
		\hInductive{\Lambda(\Var)}(X) &= (\forall x \in \Var. \exists p \in X. p = (0,x)) \\
		&\land (\forall x \in \Var. \forall M \in X. \exists p \in X. p = (1,(x,M))) \\
		& \land (\forall M \in X. \forall N \in X. \exists p \in X. p = (2, (M,N))).
	\end{align*}
Strictly speaking this is not quite a $\Delta_0$ formula, but it is not difficult to see that the statements $p = (0,x)$, $p = (1,(x,M))$ and $p = (2,(M,N))$ are expressible as $\Delta_0$ formulas (with the usual set-theoretic representations of ordered pairs). Then, we define:
	\[
	\hInductive{\Lambda(D,\Var,\Const)}(X) = \hInductive{\Lambda(\Var)}(X) \land (\forall d \in \Const. \exists p \in X. p = (3,d))
	\]
	Then, $\Lambda(\Var)$ and $\Lambda(D,\Var,\Const)$ are the least sets satisfying their respective formulas. 
\end{exa}

\begin{prop}
	\label{LambdaTermInternalProp}
	Let $\Var$ be a set\footnote{Not Boolean-valued, just the usual kind of set.}. For any complete Boolean algebra $A$,  there exists\footnote{By Theorem \ref{SetTheoryMetaTheorem} (i) and (iii).} $\Lambda(\verywidecheck{\Var})^A$ that satisfies the definition of $\Lambda(\verywidecheck{\Var})$ in the Boolean-valued sense, \emph{i.e.} $$\| \hInductive{(\Lambda(\verywidecheck{\Var})}(\Lambda(\verywidecheck{\Var})^A) \| = 1$$ and for all $X \in V^A$ such that $\| \hInductive{\Lambda(\verywidecheck{\Var})}(X) \| = 1$, we have $\| \Lambda(\verywidecheck{\Var})^A \subseteq X \| = 1$. 
	\\Then $\| \verywidecheck{\Lambda(\Var)} = \Lambda(\verywidecheck{\Var})^A \| = 1$.
\end{prop}
\begin{proof}
We know that $\hInductive{\Lambda(\Var)}$ holds of $\Lambda(\Var)$, so by Theorem \ref{DeltaNoughtMetaTheorem}, \\ $\| \hInductive{\Lambda(\verywidecheck{\Var})}(\verywidecheck{\Lambda(\Var)}) \| = 1$. By the defining property of $\Lambda(\verywidecheck{\Var})^A$, we have $\| \Lambda(\verywidecheck{\Var})^A \subseteq \verywidecheck{\Lambda(\Var)} \| = 1$, and we only need to show the opposite inclusion. 
	
We do this by showing that if $X \in V^A$ such that $\| \hInductive{\Lambda(\verywidecheck{\Var})}(X) \| = 1$, then $\| \verywidecheck{\Lambda(\Var)} \subseteq X \| = 1$. By expanding the definitions, $\| \verywidecheck{\Lambda(\Var)} \subseteq X \| = 1$ is equivalent to $\forall M \in \Lambda(\Var). \| \check{M} \in X \| = 1$, so we show that this follows from $\| \hInductive{\Lambda(\verywidecheck{\Var})}(X) \| = 1$ by induction on the structure of elements of $\Lambda(\Var)$. 
	
\begin{itemize}
\item Base case for a variable: 
		
If $(0,x) \in \Lambda(\Var)$ where $x \in \Var$, we start with the observation that $\verywidecheck{(0,x)} = (0^A,\check{x})^A$. Since $\|\hInductive{\Lambda(\verywidecheck{\Var})}(X) \| = 1$ we have $\| (0^A,\check{x})^A \in X \| = 1$, and therefore $\verywidecheck{(0,x)} \in X \| = 1$, as required. 
\item Inductive step for a $\lambda$-abstraction:
		
If $(1,(x,M)) \in \Lambda(\Var)$ such that $x \in \Var$ and $M \in \Lambda(\Var)$ and $\| \check{M} \in X \| = 1$, then again we start with the observation that $\verywidecheck{(1,(x,M))} = (1^A,(\check{x},\check{M})^A)^A$. We can deduce from $\| \hInductive{\Lambda(\verywidecheck{\Var})}(X) = 1 \|$ that $\| (1^A,(\check{x},\check{M})^A)^A \in X \| = 1$, and therefore $\| \verywidecheck{(1,(x,M))} \in X \| = 1$. 
		
\item Inductive step for an application:
		
If $(2,(M,N)) \in \Lambda(\Var)$ such that $M,N \in \Lambda(\Var)$ and $\| \check{M} \in X \| = \| \check{N} \in X \| = 1$, then as in the previous cases, we start with $\verywidecheck{(2,(M,N))} = (2^A,(\check{M},\check{N})^A)^A$. As $\| \hInductive{\Lambda(\verywidecheck{\Var})}(X)\|  = 1$ we have $\| (2^A,(\check{M},\check{N})^A)^A \in X \| = 1$, so $\| \verywidecheck{(2, (M,N))} \in X \| = 1$. 
\end{itemize}
This completes the induction. 
\end{proof}

The equational theory of $\lambda$-calculus, called $\lamth$ is described following \cite{barendregt84}:

\begin{defi}
	\label{LamthDef}
	Sentences of $\lamth$ are of the form $M = N$, where $M,N \in \Lambda(\Var)$. The theory $\lamth$ is generated by the following rules, under modus ponens and $\alpha$-conversion. 
	\begin{enumerate}[(i)]
		\item $(\lambda x. M) N = M[x := N]$, where $M[x := N]$ is capture-avoiding substitution of $N$ for $x$. 
		\item $M = M$.
		\item $M = N \Rightarrow N = M$.
		\item $M = N, N = L \Rightarrow M = L$.
		\item $M = N \Rightarrow MZ = NZ$.
		\item $M = N \Rightarrow ZM = ZN$.
		\item $M = N \Rightarrow \lambda x. M = \lambda x. N$. \qedhere
	\end{enumerate}
\end{defi}
We write $\lamth \vdash M = N$ to say that the equation $M = N$ is derivable in the theory $\lamth$. 

\begin{exa}
	\label{LamthSetExample}
	Here is a brief description of the set-theoretic formulation of $\lamth$. We can simply consider it to be a subset of $\Lambda(\Var) \times \Lambda(\Var)$, and interpret $\lamth \vdash M = N$ as $(M,N) \in \lamth$. We can define a formula $\hInductive{\lamth}(S)$, which we no longer require to be $\Delta_0$, that encodes the rules of forming $\lamth$ from Definition \ref{LamthDef}, and then $\lamth$ can be characterized as the least set (with respect to $\subseteq$) such that $\hInductive{\lamth}(\lamth)$. This allows us to define $\lamth^A$ in $V^A$. 
	
	We can then prove that $\| \check{\lamth} \subseteq \lamth^A\| = 1$ by induction on the structure of an element of $\lamth$. 
\end{exa}


\begin{defi}
	\label{LambdaInterpDef}
	Following \cite[Definition 5.4.2]{barendregt84}, we define the interpretation of $\lambda$-terms from $M \in \Lambda(D,\Var,\Const)$ in $D$, using a valuation $\rho$ such that $\fv(M) \subseteq \dom(\rho)$, by
	\begin{align*}
		\sembrack{x}_\rho &= \rho(x) & x \in \Var \\
		\sembrack{k}_\rho &= k & k \in \Const \\
		\sembrack{MN}_\rho &= \sembrack{M}_\rho \cdot \sembrack{N}_\rho \\
		\sembrack{\lambda x. M}_\rho &= \lam(\bblambda d. \sembrack{M}_{\rho(x := d)}),
	\end{align*}
	where $\bblambda$ is the ``meta-lambda'', and $\rho(x := d)$ is the partial function where $\dom(\rho(x:=d)) = \dom(\rho) \cup \{x\}$ and
	\[
	\rho(x:=d)(y) = \begin{cases} \rho(y) & \text{if } y \in \dom(\rho) \setminus \{x\} \\
		d &\text{if } y = x \end{cases}
	\]
	This is proved to define a model of $\lambda$-calculus in \cite[Theorem 5.4.4]{barendregt84}, which is to say, that for all $M, N \in \Lambda(D,\Var)$, if $\lamth \vdash M = N$, then for all valuations $\rho$, $\sembrack{M}_\rho = \sembrack{N}_\rho$. For closed terms $M \in \Lambda(D,\Var,\Const)$, we write $\sembrack{M}$ for $\sembrack{M}_\emptyset$. \qedhere
\end{defi}

We now give the Boolean-valued version. 
\begin{thm}
\label{BVSemanticsTheorem}
Let $D \in V^A$ such that the $A$-setoid of $D$ is strict, total and complete, and let there be $\lam$ and $\cdot$ in $V^A$ such that $\| (D,\lam,\cdot) \text{ is a reflexive dcpo} \| = 1$. And take $\Var$ to be a set of variables and $\Const \in V^A$ such that $\| \Const \subseteq D \| = 1$ an $A$-set of constants. Then for each $\rho$ that defines a partial function from $\verywidecheck{\Var}$ to $D$, there are functions $\sembrack{\blank}^A_\rho \in \SetoidF_A(\Lambda(D,\verywidecheck{\Var},\Const)^A,D)$ and $\sembrack{\blank}^A_\rho \in \SetoidF_A(\verywidecheck{\Lambda(\Var)},D)$ satisfying Definition \ref{LambdaInterpDef} in the $A$-valued sense. 
	
Furthermore, for all $M,N \in \Lambda(\Var)$, if $\lamth \vdash M = N$, then for all $\rho$ we have $\sembrack{\check{M}}_\rho = \sembrack{\check{N}}_\rho$. 
\end{thm}
\begin{proof}
It follows from Theorem \ref{SetTheoryMetaTheorem} applied to Definition \ref{LambdaInterpDef} that there exist some elements of $V^A$ representing functions $\sembrack{\blank}^A_\rho$ with the required properties holding with Boolean value $1$. So by Corollary \ref{RelFuncExistenceCor} we can take $\sembrack{\blank}^A_\rho \in \Set_A(\Lambda(D,\verywidecheck{\Var},\Const)^A,D)$ and also $\sembrack{\blank}^A_\rho \in \Set_A(\verywidecheck{\Lambda(\Var)},D)$ for pure terms (using Proposition \ref{LambdaTermInternalProp}). So by Definition \ref{OidFunctorDef}, these define maps in $\SetoidR_A(\Lambda(D,\verywidecheck{\Var},\Const)^A,D)$ and $\SetoidR_A(\verywidecheck{\Lambda(\Var)},D)$. So by completeness of $D$, we can apply Definition \ref{CeilFDef} to get maps in $\SetoidF_A(\Lambda(D,\verywidecheck{\Var},\Const)^A,D)$ and $\SetoidF_A(\verywidecheck{\Lambda(\Var)},D)$. 
	
From this, we get that if $\lamth \vdash M = N$, then $\| \verywidecheck{(M,N)} \in \lamth^A \| = 1$, and therefore for all $\rho$, $\| \sembrack{\check{M}}_\rho = \sembrack{\check{N}}_\rho \| = 1$. Since $D$ is strict, this implies that actually $\sembrack{\check{M}}_\rho = \sembrack{\check{N}}_\rho$ in the usual sense as well. 
\end{proof}

\subsection{The Engeler Model}
We start with a basic result about the power set. 

\begin{prop}
	\label{PowerSetDomainProp}
	For any set $X$, the power set $\powerset(X)$, when ordered using the subset relation $\subseteq$, is a dcpo (in fact, a complete lattice). If $S,T \in \powerset(X)$, $S \ll T$ iff $S$ is finite and $S \subseteq T$. The set of finite sets $\pfin(X)$ is a base, and if $X$ is countable, it is a countable base. 
\end{prop}
We can then apply this in $V^A$ as follows. 

\begin{prop}
	\label{PowerSetBVDomainProp}
	For any $X \in V^A$, the $A$-set $\apowerset(X)$, ordered with the subset relation, is a continuous lattice with base $\apfin(X)$, interpreted in the $A$-valued sense. As an $A$-setoid, $\apowerset(X)$ is complete and total. If $X = \check{Y}$, then $\verywidecheck{\pfin(Y)}$ is a base, and $\apowerset(\check{Y})$ is strict. 
\end{prop}
\begin{proof}
We get the first part by applying Theorem \ref{SetTheoryMetaTheorem} to Proposition \ref{PowerSetDomainProp}. By Theorem \ref{PowerSetCompletenessTheorem}, $\apowerset(X)$ is complete, and it is clear from its definition that it is total. Now, if $X = \check{Y}$, we use the fact that $\| \verywidecheck{\pfin(Y)} = \apfin(\check{Y}) \| = 1$ (Proposition \ref{FiniteSetPreservationProp}), so $\verywidecheck{\pfin(Y)}$ is also a base for $\apowerset(\check{Y})$ with Boolean value 1. 
	
Finally, to show that $\apowerset(\check{Y})$ is strict, let $S,T \in \apowerset(\check{Y})$ such that $\| S = T \|_{\apowerset(\check{Y})} = 1$, which is equivalent to $\| S = T \| = 1$. First observe that $\dom(S) = \dom(\check{Y}) = \dom(T)$, and then that for all $\check{y} \in \dom(\check{Y})$, we have $\| \check{y} \in S \| = S(\check{y})$ and likewise for $T$. Since $\| S = T \| = 1$, for all $y \in Y$ we have 
\[
	S(\check{y}) = \| \check{y} \in S \| = \| \check{y} \in T \| = T(\check{y}),
\]
so $S = T$. 
\end{proof}

We define the Engeler model \cite{engeler1981,barendregt84}, an adaptation of the Scott-Plotkin graph model that uses set-theoretic operations instead of G\"odel numbering, as follows\footnote{What we really need of $E$ is for it to be countable, non-empty, and an algebra of the functor $\pfin \times \Id$, such that the structure map $\pfin(E) \times E \rightarrow E$ is injective, which implies $E$ must be infinite. We cannot use the initial algebra of $\pfin \times \Id$, because it is $\emptyset$.}. 

Let $E_0 = \{ \emptyset \}$. We define $E_{n+1} = (\pfin(E_n) \times E_n) \uplus E_n$ and $E = \bigcup_{n=0}^\infty E_n$. Then we have a map $( \blank, \blank ) : \pfin(E) \times E \rightarrow E$ defined by pairing, since for each $S \in \pfin(E)$ we have $S \subseteq E_n$ for some $n \in \N$. For reasons that will become clear later, we express the existence of the Engeler model in the following manner.

\begin{prop}
	\label{EngelerExistenceProp}
	Let $E$ be a set equipped with an injective mapping $(\blank,\blank) : \pfin(E) \times E \rightarrow E$. Define $\lam : \scottcon{\powerset(E)}{\powerset(E)} \rightarrow \powerset(E)$ and $\blank \cdot \blank : \powerset(E) \times \powerset(E) \rightarrow \powerset(E)$ by
	\begin{align*}
		\lam(f) &= \{ (K,q) \mid q \in f(K) \} = \{ x \in E \mid \exists K \in \pfin(E), q \in f(K) . x = (K,q) \} \\
		F \cdot X &= \{ q \in E \mid \exists K \in \pfin(E) . K \subseteq X \text{ and } (K,q) \in F \},
	\end{align*}
	where $f \in \scottcon{\powerset(E)}{\powerset(E)}$, and $F,X \in \powerset(E)$. Then $(\powerset(E), \cdot, \lam)$ is a reflexive dcpo. 
\end{prop}

We can now build the Engeler model in $V^A$, as follows. 
\begin{thm}
\label{EngelerModelTheorem}
Let $E$ be the set defined before Proposition \ref{EngelerExistenceProp}. $(\apowerset(\check{E}), \|\subseteq\|, \cdot, \lam)$ is an $A$-valued reflexive dcpo, where $\cdot$ and $\lam$ are defined by the $A$-valued interpretation of the definitions given in Proposition \ref{EngelerExistenceProp}. 
\end{thm}
\begin{proof}
By Theorem \ref{DeltaNoughtMetaTheorem}, $\verywidecheck{(\blank,\blank)}$ defines an injective function $\verywidecheck{\pfin(E)} \atimes \check{E} = \verywidecheck{\pfin(E) \times E} \rightarrow \check{E}$. Since $\| \verywidecheck{\pfin(E)} = \apfin(\check{E}) \| = 1$ by Prop. \ref{FiniteSetPreservationProp}, it also defines an injective function $\apfin(\check{E}) \atimes \check{E} \rightarrow \check{E}$. We apply Theorem \ref{SetTheoryMetaTheorem} (i) to Prop. \ref{EngelerExistenceProp} to conclude that $\apowerset(\check{E})$ is a reflexive dcpo.
\end{proof}

By combining Theorem \ref{EngelerModelTheorem} with Proposition \ref{PowerSetBVDomainProp} and Theorem \ref{BVSemanticsTheorem}, we obtain a function $\sembrack{\blank}^A$ that interprets $\lambda$-terms in a manner that respects provable equations between them. 

The following lemma shows that if we only use pure $\lambda$-terms we do not get anything new by using $\apowerset(\check{E})$ instead of $\powerset(E)$ as a model of $\lambda$-calculus. 
\begin{lem}
	\label{ClosedImpliesClassicalEngelerLemma}
	Let $M \in \Lambda(\Var)$ and $\rho$ a valuation such that $\fv(M) \subseteq \dom(\rho)$, and consider $\sembrack{\blank}^A$ and $\sembrack{\blank}$ as defined for $\apowerset(\check{E})$ and $\powerset(E)$ respectively. Then
	$
	\| \sembrack{\check{M}}^A_{\check{\rho}} = \verywidecheck{\sembrack{M}_\rho} \| = 1.
	$
	\\In particular, since $\emptyset = \check{\emptyset}$, if $M$ is closed we have
	$
	\| \sembrack{\check{M}}^A = \verywidecheck{\sembrack{M}} \| = 1.
	$
\end{lem}
\begin{proof}
We prove it by induction on the structure of $M$. The inductive hypothesis is for the statement to hold for \emph{all} valuations $\rho$, we do not do a separate induction proof for each valuation. 
\begin{itemize}
\item Base case $M = x$ with $x \in \Var$:

By definition, $\sembrack{\check{x}}^A_{\check{\rho}} = \check{\rho}(\check{x})$, by which we really mean $\| (\check{x},\sembrack{\check{x}}^A_{\check{\rho}})^A \in \check{\rho} \| = 1$. We also have $\| \verywidecheck{(x,\sembrack{x}_\rho)} \in \check{\rho}\| = 1$ and $(\check{x},\verywidecheck{\sembrack{x}_\rho})^A = \verywidecheck{(x,\sembrack{x}_\rho)}$, so since $\check{\rho}$ is a partial function internally in $V^A$, we can conclude that $\| \sembrack{\check{x}}^A_{\check{\rho}} = \verywidecheck{\sembrack{x}_\rho} \| = 1$. 
\item Inductive step for an application:
		
We first prove the key step, which is that if $S,T \subseteq E$, then
\begin{equation}
\label{EngelerApplStepEqn}
\| \check{S} \cdot \check{T} = \verywidecheck{S \cdot T} \| = 1.
\end{equation}
For all $q \in E$, we have
\begin{align*}
\| \check{q} \in \check{S} \cdot \check{T} \| &= \| \exists K \in \apfin(\check{E}) . K \subseteq \check{T} \text{ and } (K,\check{q})^A \in \check{S} \| \\
&= \| \exists K \in \verywidecheck{\pfin(E)} . K \subseteq \check{T} \text{ and } (K,\check{q})^A \in \check{S} \| & \text{Prop. \ref{FiniteSetPreservationProp}} \\
&= \bigvee_{K \in \pfin(E)} \| \check{K} \subseteq \check{T} \| \land \| (\check{K},\check{q})^A \in \check{S} \| \\
 &= \bigvee_{K \in \pfin(E)} \| \check{K} \subseteq \check{T} \| \land \| \verywidecheck{(K,q)} \in \check{S} \| \\
&= \iver{\exists K \in \pfin(E) . K \subseteq T \text{ and } (K,q) \in S}_A \\
&= \iver{q \in S \cdot T}_A  = \| \check{q} \in \verywidecheck{S \cdot T} \|,
\end{align*}
where we have used the $A$-valued Iverson bracket $\iver{\blank}_A$ above. 
		
Then the proof of this step proceeds as follows, where for convenience we write at each step simply $A = B$ instead of $\| A = B \| = 1$:
\begin{align*}
\sembrack{\verywidecheck{MN}}^A_{\check{\rho}} &= \sembrack{\check{M}\check{N}}^A_{\rho} \\
&= \sembrack{\check{M}}^A_{\check{\rho}} \cdot \sembrack{\check{N}}^A_{\check{\rho}} \\
&= \verywidecheck{\sembrack{M}_\rho} \cdot \verywidecheck{\sembrack{N}_\rho} & \text{by inductive hypothesis} \\
&= \verywidecheck{\sembrack{M}_\rho \cdot \sembrack{N}_\rho} & \text{\eqref{EngelerApplStepEqn}} \\
&= \verywidecheck{\sembrack{MN}_\rho}. 
\end{align*}
\item Inductive step for a $\lambda$-abstraction:
		
For this step, we first point out that when interpreting $\sembrack{\lambda x . M}_\rho$ in the Engeler model, we can simplify the expression
\begin{align*}
\lam(\bblambda d. \sembrack{M}_{\rho(x := d)}) &= \{ (K,q) \in E \mid  q \in \sembrack{M}_{\rho(x := K)} \}
\end{align*}
So for $p \in E$ we have:
\begin{align*}
& \| \check{p} \in \sembrack{\lambda \check{x}. \check{M}}^A_{\check{\rho}} \| \\
={} & \| \exists K \in \apfin(\check{E}), q \in \check{E} . \check{p} = (K,q) \text{ and } q \in \sembrack{\check{M}}^A_{\check{\rho}(\check{x} := K)} \| \\
={} & \| \exists \check{K} \in \verywidecheck{\pfin(E)}, \check{q} \in \check{E} . \check{p} = \verywidecheck{(K,q)} \text{ and } \check{q} \in \sembrack{\check{M}}^A_{\check{\rho}(\check{x} := \check{K})} \| & \text{Proposition \ref{FiniteSetPreservationProp}} \\
={} & \| \exists \check{K} \in \verywidecheck{\pfin(E)}, \check{q} \in \check{E} . \check{p} = \verywidecheck{(K,q)} \text{ and } \check{q} \in \sembrack{\check{M}}^A_{\scalebox{0.7}{\verywidecheck{\rho(x := K)}}} \| \\
={} & \| \exists \check{K} \in \verywidecheck{\pfin(E)}, \check{q} \in \check{E} . \check{p} = \verywidecheck{(K,q)} \text{ and } \check{q} \in \verywidecheck{\sembrack{M}_{\rho(x := K)}} \| &\text{by inductive hypothesis} \\
={} & \iver{p \in \sembrack{\lambda x. M}_\rho}_A  ={} \| \check{p} \in \verywidecheck{\sembrack{\lambda x. M}_\rho} \|.
\end{align*}
\end{itemize}
\end{proof}
	
\subsection{Injective Spaces and Oracles}
We can encode the Booleans $\{\bot,\top\}$, and natural numbers $\N$ in $\lambda$-calculus in several ways, e.g. with the Church numerals. The details of the encoding do not matter, only the following.

\begin{defi}
	\label{EncodingDef}
	Let $(D, \fun, \lam)$ be a reflexive dcpo, and let $\bot, \top$ be closed $\lambda$-terms and $(c_n)_{n \in \N}$ a family of closed $\lambda$-terms. We say $(D,\fun,\lam,\bot,\top,(c_n)_{n \in \N})$ is a \emph{reflexive dcpo with numerals} if
	\begin{enumerate}[(i)]
		\item $\sembrack{\bot}$ and $\sembrack{\top}$ are distinct elements of $D$.
		\item There exists a closed $\lambda$-term $\If$ such that for all $\lambda$-terms $M,N$, $\lamth \vdash \If \top M N = M$ and $\lamth \vdash \If \bot M N = N$. 
		\item There exist closed $\lambda$-terms $\succ$ and $\pred$ representing the successor and predecessor operations on $(c_n)_{n \in \N}$, \emph{i.e.} for all $n \in \N$, $\lamth \vdash \succ c_n = c_{n+1}$ and $\lamth \vdash \pred c_{n+1} = c_n$ and $\lamth \vdash \pred c_0 = c_0$. 
		\item There exists a closed $\lambda$-term $\Zero$ such that $\lamth \vdash \Zero c_0 = \top$ and for all $n > 0$, $\lamth \vdash \Zero c_n = \bot$. 
	\end{enumerate}
\end{defi}

\begin{prop}
	\label{GraphEngelerHaveNumeralsProp}
	The Church Booleans and Church numerals make the Engeler model into a reflexive continuous lattice with numerals.
\end{prop}

\begin{cor}
	\label{EngelerBVNumerals}
	The Engeler model in $V^A$, as described in Theorem \ref{EngelerModelTheorem}, is a reflexive continuous lattice with numerals, when given ($\check{\blank}$ of) the Church Booleans and Church numerals, with this whole statement being interpreted in $V^A$. 
\end{cor}
\begin{proof}
By Lemma \ref{ClosedImpliesClassicalEngelerLemma}, $\| \sembrack{\check{\bot}}^A = \verywidecheck{\sembrack{\bot}} \| = 1$, and likewise for $\top$ and the Church numerals. 
	
Then part (i) of Definition \ref{EncodingDef} is a $\Delta_0$ statement, so it follows by applying Theorem \ref{DeltaNoughtMetaTheorem} to Proposition \ref{GraphEngelerHaveNumeralsProp}. Parts (ii) - (iv) then follow by Example \ref{LamthSetExample}. 
\end{proof}
In the following, for a set $A \subseteq \N$, we write $\chi_A$ for the function $A \rightarrow 2$ such that
$
\chi_A(x) = \begin{cases} 1 & \text{if } x \in A \\
	0 & \text{if } x \not\in A \end{cases},
$
\\while if $\{ \bot,\top \}$ are part of the structure of a reflexive dcpo with numerals $D$, we write $\chi_A^D$ for the function $\N \rightarrow \{ \sembrack{\bot}, \sembrack{\top}\}$ such that:
\\$
\chi_A^D(x) = \begin{cases} \sembrack{\top} & \text{if } x \in A \\
	\sembrack{\bot} & \text{if } x \not\in A \end{cases}.
$

Every dcpo is a $T_0$ topological space when equipped with its Scott topology \cite{gierz}. 

And  a $T_0$ topological space $X$ is injective for subspace embeddings in the category of $T_0$-spaces iff $X$ is a continuous lattice equipped with the Scott topology \cite{scott72}. This has useful consequences for representations of the natural numbers in reflexive dcpos that are continuous lattices. 

\begin{lem}
	\label{EncodingUsefulLemma}
	Let $(D,\fun,\lam,\bot,\top,(c_n)_{n \in \N})$ be a reflexive dcpo with numerals. Then:
	\begin{enumerate}[(i)]
		\item $\{ \sembrack{\bot}, \sembrack{\top} \}$ and $\{ \sembrack{c_n} \}_{n \in \N}$ are discrete in the Scott topology of $D$, and also are distinct elements, \emph{i.e.} $\sembrack{c_m} = \sembrack{c_n}$ implies $m = n$. 
	\end{enumerate}
	If $D$ is additionally a continuous lattice, then:
	\begin{enumerate}[(i)]
		\setcounter{enumi}{1}
		\item For every set $A \subseteq \N$, there is a $d_g \in D$ such that for all $n \in \N$, $d_g \cdot \sembrack{c_n} = \chi^D_A(n)$. 
	\end{enumerate}
\end{lem}
\begin{proof}
\begin{enumerate}[(i)]
\item As Scott topologies are $T_0$, part (i) of Definition \ref{EncodingDef} implies that there is a Scott-open set $U \subseteq D$ containing one of $\{ \sembrack{\top}, \sembrack{\bot} \}$ but not the other. We start under the assumption that $\sembrack{\top} \in U$ and $\sembrack{\bot} \not\in U$. This implies that $\{\sembrack{\top}\}$ is an open subset of the subspace $\{\sembrack{\top},\sembrack{\bot}\}$.
		
The map $\fun(\sembrack{\lnot}) : D \rightarrow D$ is Scott continuous, and $\fun(\sembrack{\lnot})(\sembrack{\bot}) = \sembrack{\top}$ and vice-versa. So $\fun(\sembrack{\lnot})^{-1}(U)$ is an open set containing $\sembrack{\bot}$, but not $\sembrack{\top}$, proving that $\{\sembrack{\bot}\}$ is an open subset of the subspace $\{\sembrack{\top},\sembrack{\bot}\}$. This proves that $\{\sembrack{\top},\sembrack{\bot}\}$ is a discrete subspace of $D$, and the proof starting with $\sembrack{\bot} \in U$ is similar.
		
To prove the discreteness and distinctness of $\{\sembrack{c_n}\}_{n \in \N}$, we will need the fact that there exist closed $\lambda$-terms $\Num{m}$ such that $\lamth \vdash \Num{m} c_n = \top$ if $m = n$ and $\lamth \vdash \Num{m} c_n = \bot$ otherwise. It is not difficult to prove directly that we can take $\Num{1} = \lambda m. \If (\Zero m) \bot (\Zero (\pred m))$, and $\Num{n} = \lambda m. \Num{n-1} (\pred m)$. So $\fun(\sembrack{\Num{n}})(\sembrack{c_m}) = \sembrack{\top}$ if $n = m$ and $\sembrack{\bot}$ otherwise. 
		
Now, if $m \neq n$ are elements of $\N$, we have $\lamth \vdash \Num{m} c_m = \top$, and $\lamth \vdash \Num{m} c_n = \bot$, so $\fun(\sembrack{\Num{m}})(\sembrack{c_m}) = \sembrack{\top} \neq \sembrack{\bot} = \fun(\sembrack{\Num{m}})(\sembrack{c_n})$. As $\fun(\sembrack{\Num{m}}$ is a function, it follows that $\sembrack{c_m} \neq \sembrack{c_n}$. 
		
To prove the discreteness of $\{\sembrack{c_n}\}_{n \in \N}$, we show that for all $m \in \N$, the singleton $\{\sembrack{c_m}\}$ is relatively open in $\{\sembrack{c_n}\}_{n \in \N}$. Let $U \subseteq D$ be a Scott-open set such that $\sembrack{\top} \in U$, but $\sembrack{\bot} \not\in U$. Then $V = \fun(\sembrack{\Num{m}})^{-1}(U)$ is a Scott-open set such that $\sembrack{m} \in V$ but $\sembrack{n} \not\in V$ for all $n \in \N$ such that $n \neq m$. 
\item Given $A \subseteq \N$, define $g : \{ \sembrack{c_n} \}_{n \in \N} \rightarrow D$ by $g(\sembrack{c_n}) = \chi^D_A(n)$. By the discreteness of $\{\sembrack{c_n}\}_{n \in \N}$, proved in the previous part, this is continuous. Since $D$ is a continuous lattice, and therefore injective \cite[Theorem 2.12]{scott72}, $g$ extends to a Scott-continuous map $\overline{g} : D \rightarrow D$, and since $(D,\lam,\fun)$ is a reflexive dcpo we can define $d_g = \lam \overline{g}$, and then for all $n \in \N$:
\[
d_g \cdot \sembrack{c_n} = \overline{g}(\sembrack{c_n}) = \chi^D_A(n).
\]
\end{enumerate}
\end{proof}

We can now define a pre-order on sets of integers to be used in the example (Theorem \ref{IncomparableDegreesTheorem}). Readers familiar with recursion theory will see that it is the $\lambda$-calculus version of comparability of many-one degrees \cite[Definition 4.8 (i), (iii)]{soare}. 
\begin{prop}
\label{ManyOneAltDefsProp}
Let $D$ be a reflexive continuous lattice with numerals. The following are equivalent for $S_1,S_2 \subseteq \N$, in which case we write $S_1 \leq_m S_2$:
\begin{enumerate}[(i)]
\item There exists a closed $\lambda$-term $M \in \Lambda(D,\Var)$ such that for all $n \in \N$, there exists $m \in \N$ such that $\lamth \vdash M c_n = c_m$, and there exist $d_{S_1}, d_{S_2} \in D$ such that for all $n \in \N$, $\sembrack{d_{S_1} c_n} = \chi_{S_1}^D(n)$, $\sembrack{d_{S_2} c_n} = \chi_{S_2}^D(n)$ and $\sembrack{d_{S_2} (M c_n)} = \sembrack{d_{S_1} c_n}$. 
\item There exists a closed $\lambda$-term $M \in \Lambda(D,\Var)$ such that for all $n \in \N$, there exists $m \in \N$ such that $\lamth \vdash M c_n = c_m$, and for all $d_{S_1},d_{S_2} \in D$ such that for all $n \in \N$, $\sembrack{d_{S_1} c_n} = \chi_{S_1}^D(n)$ and $\sembrack{d_{S_2} c_n} = \chi_{S_2}^D(n)$, we have that for all $n \in \N$, $\sembrack{d_{S_2} (M c_n)} = \sembrack{d_{S_1} c_n}$. 
\end{enumerate}
\end{prop}
\begin{proof}
\begin{itemize}
\item (i) $\Rightarrow$ (ii):
		
Suppose that (i) holds, and let $d_{S_1}',d_{S_2}' \in D$ such that for all $n \in \N$, $\sembrack{d_{S_1}' c_n} = \chi_{S_1}^D(n)$ and $\sembrack{d_{S_2}'c_n} = \chi_{S_2}^D(n)$. Then we have $\sembrack{d'_{S_1}c_n} = \sembrack{d_{S_1}c_n}$ and likewise for $S_2$. Let us write $f(n)$ for the $m \in \N$ such that $\sembrack{Mc_n} = \sembrack{c_m}$. Then we have:
\[
\sembrack{d_{S_2}'(Mc_n)} = d_{S_2}' \cdot c_{f(n)} = d_{S_2} \cdot c_{f(n)} = \sembrack{d_{S_2} (Mc_n)} = \sembrack{d_{S_1}c_n} = \sembrack{d'_{S_1}c_n},
\]
using (i) for the second-to-last step, and proving (ii). 
\item (ii) $\Rightarrow$ (i):
		
If (ii) holds, we can avoid the vacuous case of the universal quantification over $d_{S_1}$ and $d_{S_2}$ by observing that Lemma \ref{EncodingUsefulLemma} (ii) implies that there exists $d_{S_1}$ such that for all $n \in \N$, $\sembrack{d_{S_1}c_n} = \chi^D_{S_1}(n)$ and likewise for $S_2$. Therefore (i) holds.
\end{itemize}
\end{proof}



\section{$\powerset(Y)$-valued Random Variables}
\label{RandomVariablesSection}
In this section we relate the Boolean-valued Engeler model in $V^A$ from Corollary \ref{EngelerBVNumerals} to the random-variable-based Boolean-valued models of $\lambda$-calculus from \cite{Bacci18b}, and give the example application (Theorem \ref{IncomparableDegreesTheorem}). 

We start with some terminology. A \emph{negligibility space} $(X,\Sigma,\En)$ is a measurable space $(X,\Sigma)$ equipped with a $\sigma$-ideal $\En$ such that $\Sigma/\En$ is a \emph{complete} Boolean algebra, not just $\sigma$-complete. We use $A(X)$ to denote $\Sigma/\En$, as it is the \emph{complete Boolean algebra associated to $X$} or simply the \emph{algebra associated to $X$}. 

For a probability space $(X,\Sigma,\mu)$, the \textit{null ideal} $\En(\mu)$ is the $\sigma$-ideal of sets $N \in \Sigma$ with $\mu(N) = 0$, and $\Sigma/\En(\mu)$ is a complete Boolean algebra \cite[322B,322C]{fremlin3}, so $(X,\Sigma,\En(\mu))$ is a negligibility space, and $\Sigma/\En(\mu)$ is known as the \emph{measure algebra} of $(X,\Sigma,\mu)$. 

Let $Y$ be a countable set. For each $y \in Y$, let $B_y = \{ S \subseteq Y \mid y \in S \}$, which form a subbasis of open sets for the positive topology\footnote{The positive topology equals the Scott topology.} on $\powerset(Y)$. All notions of measurability for $\powerset(Y)$ will be with respect to the Borel $\sigma$-algebra defined by this topology, and since this is a second countable topology, $(B_y)_{y \in Y}$ is also a countable generating set for this $\sigma$-algebra. 

\begin{defi}
	\label{LNoughtPosetDef}
	Let $X = (X,\Sigma_X)$ be a measurable space and $Y$ a countable set. We define $\Ell^0(X;\powerset(Y))$ to be the set of measurable functions $X \rightarrow \powerset(Y)$, using the Borel $\sigma$-algebra of the Scott topology of $\powerset(Y)$. We define a $\Sigma_X$-valued order as follows:
	\[
	\| a \leq b \|_{\Ell^0} = \{ x \in X \mid a(x) \subseteq b(x) \}.
	\]
	
	The corresponding notion of equality is
	\[
	\| a = b \|_{\Ell^0} = \{ x \in X \mid a(x) = b(x) \}.
	\]
\end{defi}

\begin{lem}
\label{lem}	
If $X = (X,\Sigma_X,\En_X)$ is a negligibility space, define $L^0(X;\powerset(Y))$ to be $\Ell^0(X;\powerset(Y))$ modulo the relation $\sim$ defined by: 
$a \sim b \text{ iff } X \setminus \| a = b \|_{\Ell^0} \in \En_X$,
which is the usual ``almost everywhere'' equivalence. Then $\bleq_{L^0}$ and $\beq_{L^0}$, defined by composing the corresponding notions for $\Ell^0(X;\powerset(Y))$ with $[\blank] : \Sigma \rightarrow A(X)$, are well-defined and make $L^0(X;\powerset(Y))$ into an $A(X)$-poset. 
\end{lem}
\begin{proof}
We prove that $\| a \leq b \|_{\Ell^0}$ is measurable as follows. First, define $\gamma \subseteq \powerset(Y) \times \powerset(Y)$ to be the graph of the $\subseteq$ relation, \emph{i.e.} $\gamma = \{ (S,T) \in \powerset(Y) \times \powerset(Y) \mid S \subseteq T \}$.
	
$\| a \leq b \|_{\Ell^0} = \langle a, b \rangle^{-1}(\gamma)$, so since $\langle a, b \rangle : X \rightarrow \powerset(Y) \times \powerset(Y)$ is measurable, we only need to show that $\gamma$ is measurable with respect to the product measurable space structure. We have
\begin{align*}
\gamma &= \{ (S,T) \in \powerset(Y) \times \powerset(Y) \mid \forall y \in Y. y \in S \Rightarrow y \in T \} \\
&= \bigcap_{y \in Y} \{ (S,T) \in \powerset(Y) \times \powerset(Y) \mid y \not\in S \text{ or } y \in T \} \\
&= \bigcap_{y \in Y} \{ (S,T) \in \powerset(Y) \times \powerset(Y) \mid y \not\in S \} \cup \{ (S,T) \in \powerset(Y) \times \powerset(Y) \mid y \in T \} \\
&= \bigcap_{y \in Y} ((\powerset(Y) \setminus B_y) \times \powerset(Y)) \cup (\powerset(Y) \times B_y).
\end{align*}
Measurability then follows from the countability of $Y$. 
	
For the well-definedness of $\bleq_{L^0}$, it is clear that if $a \sim a'$ and $b \sim b'$, then $\| a \leq b \|_{\Ell^0}$ and $\| a' \leq b' \|_{\Ell^0}$ can only differ where $a$ differs from $a'$ or $b$ differs from $b'$, and the set of all such points is in $\En_X$. 

It is easy to deduce the transitivity (in the sense of Definition \ref{BVPosetDef} (i)) from transitivity of $\subseteq$, and part (ii) of the definition of an $A(X)$-poset follows because $\| a \leq a \|_{L^0} = 1$ for all $a \in L^0(X;\powerset(Y))$. 
\end{proof}

We define $G_X : L^0(X;\powerset(Y)) \rightarrow \axpowerset(\check{Y})$ as follows, where $a \in \Ell^0(X;\powerset(Y))$ and $y \in Y$:
\begin{equation}\label{GXDefEqn}	
	G_X([a])(\check{y}) = [a^{-1}(B_y)] = [\{x \in X \mid y \in a(x) \}]
\end{equation}
The measurability of $a$ guarantees that $a^{-1}(B_y) \in \Sigma$. 

\begin{prop}
\label{GXBijectionProp}
For any negligibility space $(X,\Sigma,\En)$ and any countable set $Y$, the map $G_X$ is a strict isomorphism of $A(X)$-posets $L^0(X;\powerset(Y)) \rightarrow \axpowerset(\check{Y})$.
\end{prop}
\begin{proof}
We show that for all $a,b \in \Ell^0(X;\powerset(Y))$, $\|G_X([a]) \leq G_X([b])\|_{\axpowerset(\check{Y})} = \| [a] \leq [b] \|_{L^0}$. This shows, in one go, that $G_X$ is an $A(X)$-monotone function and also injective (using the fact that $L^0(X;\powerset(Y))$ is total, \emph{i.e.} $\inn_{L^0(X;\powerset(Y))}([a]) = [X]$ for all $a \in \Ell^0(X;\powerset(Y))$). 
	
We start by expanding the definitions:
\begin{align*}
& \| G_X([a]) \leq G_X([b]) \|_{\axpowerset(\check{Y})} = \| G_X([a]) \subseteq G_X([b]) \| \\
={} & \bigwedge_{t \in \dom(G_X([a]))} \| t \in G_X([a]) \| \Rightarrow \| t \in G_X([b]) \| = \bigwedge_{y \in Y} G_X([a])(\check{y}) \Rightarrow G_X([b])(\check{y}) \\
={} & \bigwedge_{y \in Y} [\{ x \in X \mid y \in a(x)\}] \Rightarrow [\{x \in X \mid y \in b(x) \}] \\
 ={} & \left[\bigcap_{y \in Y} \{ x \in X \mid y \in a(x) \text{ implies } y \in b(x) \}\right] \\
={} & [\{ x \in X \mid \forall y \in Y. y \in a(x) \text{ implies } y \in b(x) \}] \\
 ={} & [\{ x \in X \mid a(x) \subseteq b(x) \}] = \| [a] \leq [b] \|_{L^0}.
\end{align*}
	
Since $L^0(X;\powerset(Y))$ is a strict $A(X)$-setoid, and $\axpowerset(\check{Y})$ is total, implying that $G_X$ is an injective function. So to prove that $G_X$ is a strict isomorphism, we only need to show that $G_X$ is surjective.
	
Let $b \in \axpowerset(\check{Y})$. For each $y \in Y$, pick $S_y \in \Sigma$ such that $[S_y] = b(\check{y})$. Define a function \\$a : X \rightarrow \powerset(Y)$ by $a(x) = \{ y \in Y \mid x \in S_y \}$. For each $y \in Y$, we have
\[
a^{-1}(B_y) = \{ x \in X \mid a(x) \in B_y \} = \{ x \in X \mid y \in a(x) \} = \{ x \in X \mid x \in S_y \} = S_y \in \Sigma.
\]
Since $(B_y)_{y \in Y}$ generates the Borel $\sigma$-algebra of $\powerset(Y)$, this implies that $a$ is measurable. For all $y \in Y$, we then have
\[
G_X([a])(\check{y}) = [a^{-1}(B_y)] = [S_y] = b(\check{y}),
\]
and therefore $G_X([a]) = b$, as required to prove $G_X$ surjective. 
\end{proof}

Since $\powerset(E)$ is a reflexive dcpo and has a binary operation $\cdot$ for application, for any measurable space $X$ we can extend this pointwise to $\Ell^0(X;\powerset(E))$ by defining for $a,b \in \Ell^0(X;\powerset(E))$ and $x \in X$:
$
(a \cdot b)(x) = a(x) \cdot b(x).
$
\\For a negligibility space $X$ we can then extend this to $L^0(X;\powerset(E))$ by defining
$
[a] \cdot [b] = [a \cdot b].
$
\begin{prop}
\label{ApplicationPointwiseWellDefProp}
Let $(X,\Sigma,\En)$ be a negligibility space. Then the above definition of $\cdot : L^0(X;\powerset(E)) \times L^0(X;\powerset(E)) \rightarrow L^0(X;\powerset(E))$ is well-defined, and $G_X : L^0(X;\powerset(E)) \rightarrow \axpowerset(\check{E})$ is an ``application homomorphism'', \emph{i.e.} for all $[a],[b] \in L^0(X;\powerset(E))$,
\[
G_X([a] \cdot [b]) = G_X([a]) \cdot G_X([b]).
\]
\end{prop}
\begin{proof}
In fact, we will prove the first part by deducing it from the second part. So let $a,b \in \Ell^0(X;\powerset(E))$. Then for all $q \in E$ we have
\begin{align*}
& (G_X([a]) \cdot G_X([b]))(\check{q}) \\
={} & \| \exists K \in \axpfin(\check{E}) . K \subseteq G_X([b]) \text{ and } (K,\check{q})^{A(X)} \in G_X([a]) \| \\
={} & \| \exists K \in \verywidecheck{\pfin(E)} . K \subseteq G_X([b]) \text{ and } (K,\check{q})^{A(X)} \in G_X([a]) \| & \text{Proposition \ref{FiniteSetPreservationProp}} \\
={} & \bigvee_{K \in \pfin(E)} \| \check{K} \subseteq G_X([b]) \| \land \| \verywidecheck{(K,q)} \in G_X([a]) \| \\
={} & \bigvee_{K \in \pfin(E)} G_X([a])(\verywidecheck{(K,q)}) \land \bigwedge_{k \in K} G_X([b])(\check{k}) \\
={} & \bigvee_{K \in \pfin(E)} [\{ x \in X \mid (K,q) \in a(x) \}] \land \bigwedge_{k \in K} [\{ x \in X \mid k \in b(x) \}] \\
={} & \bigvee_{K \in \pfin(E)} [\{ x \in X \mid (K,q) \in a(x) \}] \land [\{ x \in X \mid K \subseteq b(x) \}] \\
={} & [\{x \in X \mid \exists K \in \pfin(E) . K \subseteq b(x) \text{ and } (K,q) \in a(x)\}] & \text{as $\pfin(E)$ countable} \\
={} & [\{x \in X \mid q \in a(x) \cdot b(x)\}] \\
={} & G_X([a \cdot b])(\check{q}),
\end{align*}
so all together we have proved that $G_X([a]) \cdot G_X([b]) = G_X([a \cdot b])$. In passing, we have proved that for all $q \in E$, $(a \cdot b)^{-1}(B_q)$ is measurable, and therefore that $a \cdot b \in \Ell^0(X;\powerset(E))$.
	
If $a',b' \in \Ell^0(X;\powerset(E))$ such that $[a'] = [a]$ and $[b'] = [b]$, then by Proposition \ref{GXBijectionProp} and the fact that $\axpowerset(\check{E})$ is a strict $A(X)$-setoid:
\[
G_X([a \cdot b]) = G_X([a]) \cdot G_X([b]) = G_X([a']) \cdot G_X([b']) = G_X([a' \cdot b']),
\]
and therefore $[a \cdot b] = [a' \cdot b']$, so the definition $[a] \cdot [b] = [a \cdot b]$ genuinely defines a function $L^0(X;\powerset(E)) \times L^0(X; \powerset(E)) \rightarrow L^0(X;\powerset(E))$. Putting this together with what we proved above gives us $G_X([a]) \cdot G_X([b]) = G_X([a] \cdot [b])$. 
\end{proof}

We now introduce the following notation. If $S \subseteq E$, define $K_S \in \Ell^0(X;\powerset(E))$ to be the function taking the constant value $S$. This relates to $\check{\blank}$ in the following way. 
\begin{lem}
\label{ConstantClassicalLemma}
Let $Y$ be an arbitrary countable set. For all $S \subseteq Y$ we have
\[
\| G_X([K_S]) = \check{S} \| = 1
\]
\end{lem}
\begin{proof}
It suffices to show that for all $y \in Y$, $\| \check{y} \in G_X([K_S]) \| = \| \check{y} \in \check{S} \|$. We have
\[
\| \check{y} \in G_X([K_S]) \| = G_X([K_S])(\check{y}) = [\{ x \in X \mid y \in K_S(x)\}] = [\{ x \in X \mid y \in S\}] = \| \check{y} \in \check{S} \|.
\]
\end{proof}

Recall that a subbase of clopens for the product topology of $2^\omega$ is given by the sets $(C_{n,b})_{n \in \omega, b \in 2}$, where
$
C_{n,b} = \{ f \in 2^\omega \mid f(n) = b \}.
$
\\These sets also generate the Borel $\sigma$-algebra of $2^\omega$. 

In the following, we will also have to consider the (isomorphic) product space $2^\omega \times 2^\omega$, but for which we will need to describe the subsets in more detail. So for $i \in \{1,2\}$, $n \in \omega$ and $b \in 2$ we write
$
D_{i,n,b} = \{ (f_1,f_2) \in 2^\omega \times 2^\omega \mid f_i(n) = b \}.
$
\\Then for all $i \in \{ 1, 2 \}$, $n \in \omega$ and $b \in 2$ we have $D_{i,n,b} = \pi_i^{-1}(C_{n,b})$. 

Let $X = 2^\omega \times 2^\omega$ equipped with its Borel $\sigma$-algebra as a measurable space. We define $\mu_X$ to be the usual independent fair coin measure, which is to say, it is the unique measure such that for all finitely-supported partial functions $f_1,f_2 : \omega \rightarrow 2$
\[
\mu\left(\bigcap_{n \in \dom(f_1)} D_{1,n,f_1(n)} \cap \bigcap_{n \in \dom(f_2)} D_{2,n,f_2(n)}\right) = 2^{-(|\dom(f_1)| + |\dom(f_2)|)}. 
\]
We take the null ideal of this measure as the negligible sets of $X$. Then $\pi_1,\pi_2$ are random variables in $\Ell^0(X;2^\omega)$. They define $S_1,S_2 \in \axpowerset(\check{\omega})$ as follows, taking $i \in \{1,2\}$:
\begin{equation}
	\label{SiDefEqn}
	S_i(\check{n}) = [\pi_i^{-1}(C_{n,1})] = [D_{i,n,1}].
\end{equation}
We remind the reader at this point that $\| \check{n} \in S_i \| = S_i(\check{n})$ because of the way Boolean-valued equality behaves for elements of $\check{\omega}$. 

\begin{prop}
\label{FiniteImageLemma}
For all $f : \omega \rightarrow \omega$, neither $S_1$ nor $S_2$ is the preimage of a finite $A(X)$-subset of $\check{\omega}$, \emph{i.e.} $\| \exists K \in \axpfin(\check{\omega}). \forall n \in \check{\omega}. n \in S_1 \Leftrightarrow \check{f}(n) \in K \| = 0$, and likewise for $S_2$. 

Moreover, $\| S_1 = \check{f}^{-1}(S_2) \| = 0$ and $\| S_2 = \check{f}^{-1}(S_1) \| = 0$, \emph{i.e.} these sets cannot be mapped to each other by any classical function.
\end{prop}
\begin{proof}
We only give the proof for $S_1$, as the proof for $S_2$ is essentially the same. Using Proposition \ref{FiniteSetPreservationProp}, we have
\begin{align*}
& \| \exists K \in \axpfin(\check{\omega}). \forall n \in \check{\omega}. n \in S_1 \Leftrightarrow \check{f}(n) \in K \| \\
 ={} & \| \exists K \in \verywidecheck{\pfin(\omega)}. \forall n \in \check{\omega}. n \in S_1 \Leftrightarrow \check{f}(n) \in K \| \\
={} & \bigvee_{K \in \pfin(\omega)} \bigwedge_{n \in \omega} \| \check{n} \in S_1 \| \Leftrightarrow \| \verywidecheck{f(n)} \in \check{K} \| \\
={} & \bigvee_{K \in \pfin(\omega)} \bigwedge_{n \in \omega} [D_{1,n,1}] \Leftrightarrow \iver{f(n) \in K}_{A(X)}. 
\end{align*}
This being equal to $0$ is equivalent to the statement that for all $K \in \pfin(\omega)$:
\[
\bigwedge_{n \in \omega} [D_{1,n,1}] \Leftrightarrow \iver{f(n) \in K}_{A(X)} = 0.
\]
Then
\begin{align*}
\bigwedge_{n \in \omega} [D_{1,n,1}] \Leftrightarrow \iver{f(n) \in K}_{A(X)} &= \bigwedge_{n \in \omega} [D_{1,n,\iver{f(n) \in K}_2}] = \left[ \bigcap_{n \in \omega} D_{1,n,\iver{f(n) \in K}_2} \right]
\end{align*}
Since it is an infinite intersection of distinct sets for distinct indices, the measure of the intersection above $\leq 2^{-n}$ for all $n$, and so it has measure zero, so the above element has Boolean value $0$, as required. 
	
For the second part, we only prove that $\| S_1 = \check{f}^{-1}(S_2) \| = 0$, as the proof of the other way is analogous. We start by ruling out some possibilities of what $f$ is like. Assume for a contradiction that $\| S_1 = \check{f}^{-1}(S_2) \| > 0$. If the image of $f$ is finite, then since $f^{-1}(S) = f^{-1}(S \cap \im(f))$ classically, by Theorem \ref{SetTheoryMetaTheorem} we have $\| S_1 = \check{f}^{-1}(S_2) \| = \| S_1 = \check{f}^{-1}(S_2 \cap \im(\check{f})) \|$ and that $\| S_2 \cap \im(\check{f})  \in \axpfin(\check{\omega}) \| = 1$. But this contradicts Lemma \ref{FiniteImageLemma}, so either $\| S_1 = \check{f}^{-1}(S_2) \| = 0$ after all, or the image of $f$ is infinite, so we reduce to the latter case.
	
If the image of $f$ is infinite, then by the well-orderedness of $\omega$ there exists an infinite set $I \subseteq \omega$ such that $f|_I$ is injective, and we fix such a set until the end of the proof. 
	
We now convert $\| S_1 = \check{f}^{-1}(S_2) \|$ into Boolean-valued set theory.
\begin{align*}
\| S_1 = \check{f}^{-1}(S_2) \| = \| \forall n \in \check{\omega} . n \in S_1 \Leftrightarrow \check{f}(n) \in S_2 \| = \bigwedge_{n \in \omega} \| \check{n} \in S_1 \| \Leftrightarrow \| \check{f}(\check{n}) \in S_2 \| 
\end{align*}
	
We can then analyse what $\| \check{f}(\check{n}) \in S_2 \|$ means as a formula in set theory:
\begin{align*}
\| \check{f}(\check{n}) \in S_2 \| &= \| \forall m \in \check{\omega}. (\check{n},m)^{A(X)} \in \check{f} \Rightarrow m \in S_2 \| = \bigwedge_{m \in \omega} \| \verywidecheck{(n,m)} \in \check{f} \| \Rightarrow \| m \in S_2 \| \\
&= \| \verywidecheck{f(n)} \in S_2 \|.
\end{align*}
	
So
\begin{align*}
\| S_1 = \check{f}^{-1}(S_2) \| &= \bigwedge_{n \in \omega} \| \check{n} \in S_1 \| \Leftrightarrow \| \verywidecheck{f(n)} \in S_2 \| \\
 &=  \bigwedge_{n \in \omega} S_1(\check{n}) \Leftrightarrow S_2(\verywidecheck{f(n)}) \leq \bigwedge_{n \in I} S_1(\check{n}) \Leftrightarrow S_2(\verywidecheck{f(n)}) \\
&= \bigwedge_{n \in I} [D_{1,n,1}] \Leftrightarrow [D_{2,f(n),1}] \\
 &= \left[ \bigcap_{n \in I} (D_{1,n,1} \cap D_{2,f(n),1}) \cup (D_{1,n,0} \cap D_{2,f(n),0})\right].
\end{align*}
We will show that this is equal to $0$ by showing that the set inside the brackets has measure zero. Recall the finite distributive law, that if we have a finite set $K$ and sequences of sets $(S_{n,0})_{n \in K}$ and $(S_{n,1})_{n \in K}$ then
\[
\bigcap_{n \in K} S_{n,0} \cup S_{n,1} = \bigcup_{g : K \rightarrow 2} \bigcap_{n \in K} S_{n,g(n)}.
\]
This is easily proved by induction on $|K|$. Note also that if for each $n \in K$, the sets $S_{n,0}$ and $S_{n,1}$ are disjoint, then the sets $\bigcap_{n \in K} S_{n,g(n)}$ are pairwise disjoint as $g$ ranges over $2^K$. 
	
So for all $K \in \pfin(I)$, we have
\begin{align*}
& \mu_X\left(\bigcap_{n \in K} (D_{1,n,1} \cap D_{2,f(n),1}) \cup (D_{1,n,0} \cap D_{2,f(n),0})\right) \\ 
={} &  \mu_X\left(\bigcup_{g : K \rightarrow 2} \bigcap_{n \in K} D_{1,n,g(n)} \cap D_{2,f(n),g(n)}\right) \\
={} & 2^{|K|} \mu_X\left(\bigcap_{n \in K} D_{1,n,g(n)} \cap D_{2,f(n),g(n)}\right) \\
={} &  2^{|K|} 2^{-2|K|} = 2^{-|K|},
\end{align*}
where in the second-to-last step we have used the fact that $f|_I$ is injective (and therefore $f|_K$ is too). Since we can exhaust $I$ by a countable increasing sequence of finite sets, the result follows. 
\end{proof}


\begin{thm}
\label{IncomparableDegreesTheorem}
There exist sets $T_1,T_2 \subseteq \N$ such that neither can be mapped to the other by a $\lambda$-definable function, \emph{i.e.} we neither have $T_1 \leq_m T_2$ nor $T_2 \leq_m T_1$ (recall Proposition \ref{ManyOneAltDefsProp}).
\end{thm}
\begin{proof}
We start by doing Boolean-valued reasoning about $S_1$ and $S_2$, as defined in \eqref{SiDefEqn}, taking $A = A(2^\omega \times 2^\omega)$, and use the fact that the Engeler model $\apowerset(\check{E})$ is ($A$-valuedly) a reflexive continuous lattice with numerals (Corollary \ref{EngelerBVNumerals}) and the characterization of many-one reductions from Proposition \ref{ManyOneAltDefsProp}. 
	
First, by Theorem \ref{SetTheoryMetaTheorem} applied to Lemma \ref{EncodingUsefulLemma} (ii), there exist $d_1,d_2 \in \apowerset(\check{E})$ such that for all $n \in \omega$ and $i \in \{ 1, 2\}$ we have $\| d_i \cdot \check{c_n} = \check{\top} \| = \| \check{n} \in S_i \|$ and $\| d_i \cdot \check{c_n} = \check{\bot} \| = \lnot \| \check{n} \in S_i \|$. 
	
Suppose $M \in \Lambda(\Var)$ such that for all $n \in \omega$, there exists $m \in \omega$ such that $\lamth \vdash M c_n = c_m$. In particular, there exists some function $f : \omega \rightarrow \omega$ such that for all $m \in \omega$, $\lamth \vdash M c_n = c_{f(n)}$. 
	
Now we consider $\sembrack{d_2 (\check{M} \check{c}_n)}^A$ and $\sembrack{d_1 \check{c}_n}^A$. In the following, for ease of notation, we write an equality where the fact that an $A$-valued equality is $1$ is meant:
\[
\sembrack{d_2 (\verywidecheck{M c_n})}^A  = d_2 \cdot \sembrack{\verywidecheck{M c_n}}^A = d_2 \cdot \sembrack{\verywidecheck{c_{f(n)}}}^A
\]
by Theorem \ref{BVSemanticsTheorem}. Therefore
\[
\| \sembrack{d_2 (\verywidecheck{M c_n})}^A = \sembrack{\check{\top}}^A \| = \| d_2 \cdot \sembrack{\verywidecheck{c_{f(n)}}}^A = \sembrack{\check{\top}}^A \| = \| \verywidecheck{f(n)} \in S_2 \|,
\]
and the corresponding negative statement for $\bot$. Likewise
\begin{align*}
\| \sembrack{d_1 \check{c}_n}^A = \sembrack{\check{\top}} \| &= \| \check{n} \in S_1 \|,
\end{align*}
and the corresponding statement for $\bot$. So all together
\[
\| d_2 \cdot \sembrack{\verywidecheck{M c_n}}^A = d_1 \cdot \sembrack{\check{c}_n}^A \| = \| \verywidecheck{f(n)} \in S_2 \Leftrightarrow \check{n} \in S_1 \|.
\]
It is proved in Proposition \ref{FiniteImageLemma} that $\| \forall n \in \check{\omega} . \check{f}(n) \in S_2 \Leftrightarrow n \in S_1 \| = 0$, so we can conclude that 
\[
\bigwedge_{n \in \omega} \|  d_2 \cdot \sembrack{\verywidecheck{M c_n}}^A = d_1 \cdot \sembrack{\check{c}_n}^A \| = 0
\]
for all $M \in \Lambda(\verywidecheck{\Var})$ that map numerals to numerals. 
	
We now consider how this appears in $L^0(X;\powerset(E))$, using $G_X^{-1}$. We introduce the $a_1,a_2 \in \Ell^0(X;\powerset(E))$ such that for $i \in \{1,2\}$ we have $[a_i] = G_X^{-1}(d_i)$, which we shall use in the following. By Proposition \ref{GXBijectionProp}, we have
\begin{align*}
0 &= \bigwedge_{n \in \omega} \| G_X^{-1}(d_2 \cdot \sembrack{\verywidecheck{M c_n}}^A) = G_X^{-1}(d_1 \cdot \sembrack{\check{c}_n}^A) \|_{L^0} \\
&= \bigwedge_{n \in \omega} \| G_X^{-1}(d_2) \cdot G_X^{-1}(\sembrack{\verywidecheck{M c_n}}^A) = G_X^{-1}(d_1) \cdot G_X^{-1}(\sembrack{\check{c}_n}^A) \|_{L^0} & \text{Proposition \ref{ApplicationPointwiseWellDefProp}} \\
&= \bigwedge_{n \in \omega} \| [a_2] \cdot G_X^{-1}(\verywidecheck{\sembrack{M c_n}}) = [a_1] \cdot G_X^{-1}(\verywidecheck{\sembrack{c_n}}) \|_{L^0} & \text{Lemma \ref{ClosedImpliesClassicalEngelerLemma}} \\
&= \bigwedge_{n \in \omega} \| [a_2] \cdot [K_{\sembrack{M c_n}}] = [a_1] \cdot [K_{\sembrack{c_n}}] \|_{L^0} & \text{Lemma \ref{ConstantClassicalLemma}} \\
&= \left[ \bigcap_{n \in \omega} \{ x \in X \mid a_2(x) \cdot K_{\sembrack{M c_n}}(x) = a_1(x) \cdot K_{\sembrack{c_n}}(x) \} \right] \\
&= [ \{ x \in X \mid \forall n \in \omega. a_2(x) \cdot \sembrack{M c_n} = a_1(x) \cdot \sembrack{c_n} \} ],
\end{align*}
so the set inside the square brackets has measure zero. 
	
Since (as long as $\Var$ is countable) $\Lambda(\Var)$ is countable, the set of $M \in \Lambda(\Var)$ that map numerals to numerals is also countable. Therefore the set of $x \in X$ such that for all $M \in \Lambda(\Var)$ mapping numerals to numerals we have $a_2(x) \cdot \sembrack{Mc_n} \neq a_1(x) \cdot \sembrack{c_n}$ has measure $1$. 
	
For all $i \in \{1,2\}$, we have, by the definition of $d_i$, for all $n \in \omega$ the statement $\| d_i \cdot \sembrack{c_n}^A = \sembrack{\top}^A \| \lor \| d_i \cdot \sembrack{c_n}^A = \sembrack{\bot}^A \| = 1$. By a similar argument to that used above, this means that the set of $x \in X$ such that $a_i(x) \cdot \sembrack{c_n} \in \{ \sembrack{\top}, \sembrack{\bot} \}$ has measure $1$, and by the countability of $\omega$ the set where this is true for both $i \in \{1,2\}$ and all $n \in \omega$ is also of measure $1$. 
	
Therefore the intersection of the sets defined in the previous two paragraphs has measure $1$, and so is non-empty and so there exists a point $x$ in it. We can then define $T_i = \{ n \in \omega \mid a_i(x) \cdot \sembrack{c_n} = \sembrack{\top} \}$, and we have proved, by Proposition \ref{ManyOneAltDefsProp}, that $T_1 \not\leq_m T_2$. 
	
To get the final result, we re-run the argument swapping the roles of $S_1$ and $S_2$, and take $x$ to be in the intersection of the corresponding sets of measure $1$, so that $T_2 \not\leq_m T_1$ either. 
\end{proof}

We could not have done this proof by using the ``fact'' that $L^0(X;\powerset(E))$ is a continuous dcpo, because this is not true, as stated in our last proposition. 

\begin{prop}
\label{NotContinuousProp}
Let $X = (X,\Sigma,\En)$ be a negligibility space such that $A(X)$ is not atomic, and let $Y$ be a non-empty countable set. Then $L^0(X;\powerset(Y))$ is not a continuous dcpo. 
\end{prop}
\begin{proof}
First, let $p = \bigvee \{ a \in A(X) \mid a \text{ an atom} \}$. As $A(X)$ is not atomic, $\lnot p \neq 0$, so there is a set $S \in \Sigma$ such that $S \not\in \En$ and $[S] = \lnot p$. By definition there are no atoms below $[S]$. Define
\[
b(x) = \begin{cases} Y & \text{if } x \in S \\
\emptyset & \text{otherwise,} \end{cases}
\]
where $x \in X$. This defines a measurable function $X \rightarrow \powerset(Y)$. As $S \not\in \En$, $[b] \neq [K_\emptyset]$. We show that $L^0(X;\powerset(Y))$ is not continuous by showing that $[b]$ is not the supremum of elements way below it, which will follow from the fact that the only element of $L^0(X;\powerset(Y))$ that is way below $[b]$ is $[K_\emptyset]$. 
	
If the only element below $[b]$ is $[K_\emptyset]$, then we are finished, so we reduce to the case that there is at least one $[a] \in L^0(X;\powerset(Y))$ such that $[K_\emptyset] < [a] < [b]$. Define $T = a^{-1}(\powerset(Y) \setminus \{\emptyset\})$, which is in $\Sigma$ because $a$ is measurable and $\powerset(Y) \setminus \{\emptyset\} = \bigcup_{y \in Y} B_y$ and is therefore a Borel set. We have $T \setminus S \in \En$, so we redefine $T$ to be $T \cap S$ if necessary to make $T \subseteq S$. 
	
Since there are no atoms below $[S]$, there are none below $[T]$ either, so
there exists a non-zero element $a_1 \in A(X)$ such that $a_1 \leq [T]$ and
$a_1$ has no atoms below it. We can repeat this argument to form a strictly descending sequence $(a_i)_{i \in \N}$ where for all $i \in \N$, $a_i$ has no atoms below it and $a_i \leq [T]$. We can define $b_i = a_i \setminus \bigwedge_{i = 1}^\infty a_i$, to get such a strictly decreasing sequence $(b_i)_{i \in \N}$, now with $\bigwedge_{i=1}^\infty b_i = 0$. Since $A(X) = \Sigma/\En$, we can find a sequence $(V_i)_{i \in \N}$ of elements of $\Sigma$ such that for all $i \in \N$, $[V_i] = b_i$, and by adjusting negligible sets it can be arranged that $(V_i)_{i \in \N}$ is also strictly descending. Defining $T_i = T \setminus U_i$, then $(T_i)_{i  \in \N}$ is strictly increasing with respect to $\subseteq$, maps to a strictly increasing sequence under $[\blank]$, and $T \setminus \bigcup_{i = 1}^\infty T_i \in \En$. Define, for
each $i \in \N$, $c_i : X \rightarrow \powerset(Y)$ as follows: 
\[
c_i(x) = \begin{cases} Y   & \text{if } x \in T_i \cup S \setminus T \\
\emptyset & \text{if } x \in T \setminus T_i \cup X \setminus S, \end{cases}
\]
where $x \in X$. Each $c_i$ is measurable for the same reasons
that $b$ is, and $([c_i])_{i \in \N}$ is a strictly increasing sequence in $L^0(X;\powerset(Y))$. Since for all $i \in \N$, $c_i(x) = \emptyset$ for all $x \in T\setminus T_i$ and $T \setminus T_i \not\in \En$, while $a(x) \neq \emptyset$
for all $x \in T \supseteq T \setminus T_i$, we have $[a] \not\leq [c_i]$. But
$\bigvee_{i = 1}^\infty [c_i] = \left[\bigvee_{i=1}^\infty c_i\right] = [b]$
because $T \setminus \bigcup_{i=1}^\infty T_i \in \En$. Therefore $a$ is not way
below $b$.  
\end{proof}

\section{Domain-Valued Random Variables More Generally}
\label{NewRandomVariablesSection}
In this section, we show that if $D$ is a continuous dcpo with a countable base, then $L^0(X;D)$ can be equipped with the structure of an $A$-poset (Definition \ref{BVPosetDef}) that is compatible with a suitable notion in $V^A$. This is similar to Scott's proof that the Dedekind reals in $V^A$ are isomorphic to $L^0(X;\R)$ (see \cite[\S 4.3]{solovay71} or \cite[Theorem 7.1]{bell}), but for a countably based domain $D$ instead of $\R$. 

We start by recalling some more background in domain theory, so we can carry over results from $V$ to $V^A$. The following is an adaptation of \cite[Definition III-4.15]{gierz}.

\begin{defi}[Abstract Base]
\label{AbstractBaseDef}
An \emph{abstract base} is a pair $(B,\prec)$ where ${\prec} \subseteq B \times B$ such that
\begin{enumerate}[(i)]
\item $B$ is non-empty. 
\item $\prec$ is transitive
\item For each $d \in B$, there exists $b \prec d$
\item For each $b_1,b_2,d \in B$ such that $b_1,b_2 \prec d$, there exists $c \in B$ such that $b_1,b_2 \prec c \prec d$. 
\item If $b_1 \neq b_2$, then there exists $c \in B$ such that either $c \prec b_1$ and $c \not\prec b_2$ or $c \prec b_2$ and $c \not\prec b_1$. 
\end{enumerate}
\emph{Important: $(B, \prec)$ has no obligation to be a poset, there is no reflexivity in general.}\footnote{In the case that $\prec$ is reflexive, it will eventually be clear that this corresponds to a base of compact elements defining an algebraic dcpo.} \qedhere
\end{defi}

\begin{prop}
\label{AbstractBaseInVAProp}
If $(B,\prec)$ is an abstract base, $\verywidecheck{(B,\prec)} = (\check{B},\verywidecheck{\prec})^A$ is an abstract base in $V^A$. 
\end{prop}
\begin{proof}
This follows directly from Theorem \ref{DeltaNoughtMetaTheorem}, after verifying that (i) - (v) of Definition \ref{AbstractBaseDef} are $\Delta_0$. 
\end{proof}

We summarize the following results about abstract bases, which are relatively easy to prove independently (they form the exercises in \cite[Exercises III-4.16 and 17]{gierz} following Definition III-4.15), so the proof of the non-definitional parts of the following definition, as well as the proof of the following proposition, are omitted. 

\begin{defi}[Rounded Ideals]
\label{RoundedIdealDef}
Let $(B,\prec)$ be an abstract base. A subset $I \subseteq B$ is called a \emph{rounded ideal} iff it is $\prec$-directed and a $\prec$-down set, \emph{i.e.}
\begin{enumerate}[(i)]
\item $I$ is not empty.
\item For each $b_1,b_2 \in I$ there exists $b \in I$ with $b_1,b_2 \prec b$. 
\item If $a \in B$ and $b \in I$ and $a \prec b$, then $a \in I$. 
\end{enumerate}
We write $\RI(B)$ for the set of rounded ideals, considered as a poset under $\subseteq$. If $(I_j)_{j \in J}$ is a monotone net of rounded ideals, then $\bigcup_{i \in I}I_j$ is a rounded ideal, so rounded ideals form a dcpo. \qedhere
\end{defi}

\begin{prop}
\label{AbstractBaseIsBaseProp}
Let $(B,\prec)$ be an abstract base. For $b \in B$, define $I_b = \{ a \in B \mid a \prec b \}$. Then $I_b$ is a rounded ideal of $B$, and the map $b \mapsto I_b$ is injective. 

For rounded ideals $I,J \subseteq B$, we have $I \ll J$ iff there exist $a,b \in B$ such that $a \prec b$ and $I \subseteq I_a$ and $I_b \subseteq J$ and $\{I_b \mid b \in B\}$ is a base for $\RI(B)$, so $\RI(B)$ is always a continuous dcpo. 

If $D$ is a non-empty dcpo and $B \subseteq D$ is a base, then $(B, \ll)$ is an abstract base. For each $d \in D$, the set $I_d \defeq \dd d \cap B$ is a rounded ideal. The mapping $d \mapsto I_d$ is a poset isomorphism $D \isoarrow \RI(B)$ with inverse $I \mapsto \bigvee I$. \qedhere
\end{prop}

When we build $\RI^A(\check{B})$, \emph{i.e.} $\RI(B)$ interpreted according to $V^A$, we would prefer to deal with elements of $S \in \apowerset(\check{B})$ such that $\bvs{S \text{ a rounded ideal}} = 1$. This is not the case with the usual way of using the axiom of separation with $\apowerset(\check{B})$. So we prove the following lemma.

\begin{lem}
\label{DefeatingEmptySetLemma}
Let $X \in V^A$, and let $P : \dom(\apowerset(X)) \rightarrow A$ be a mapping that is well-defined with respect to $\beq$ in the sense that for all $S_1,S_2 \in \dom(\apowerset(X))$
\[
P(S_1) \land \bvs{S_1 = S_2} \leq P(S_2)
\]
(by symmetry, we could put $=$ instead of $\leq$). We will use this in the case that $P = \bvs{\Phi(x)}$ where $\Phi$ is a formula of set theory with one free variable $x$, considered as a partial function restricted to $\dom(\powerset(X))$. 

We define two elements of $V^A$, written $E_P$ and $F_P$ as follows.
\begin{align*}
\dom(E_P) &= \{ S : \dom(X) \rightarrow A \mid P(S) = 1 \} \\ 
E_P(S) &= 1 \\
\dom(F_P) &= \{ S : \dom(X) \rightarrow A \} = \dom(\apowerset(X)) \\
F_P(S) &= P(S)
\end{align*}
Then $F_P$ is obtained from the usual proof of the axiom of separation in $V^A$ applied to $\apowerset(X)$, and $\bvs{E_P \subseteq F_P} = 1$. If there exists $T \in \dom(E_P)$, \emph{i.e.} $\dom(E_P)$ is not empty, then $\bvs{F_P \subseteq E_P} = 1$, so $\bvs{E_P = F_P} = 1$. The existence of $T$ also implies $\Oid(E_P)$ is complete. 
\end{lem}
\begin{proof}
We have $\bvs{E_P \subseteq F_P} = 1$ iff for all $S \in \dom{E_P}$, $E_P(S) \leq \bvs{S \in F_P}$. By definition $E_P(S) = 1$, and by the proof of the axiom of separation in $V^A$ (see \emph{e.g.} \cite[Theorem 14.24]{jech}), $\bvs{S \in F_P} = P(S) = 1$, since $S \in \dom{E_P}$. Therefore we conclude via $1 \leq 1$. 

Now suppose that there exists $T \in \dom{E_P}$, \emph{i.e.} $T : \dom(X) \rightarrow X$ and $P(T) = 1$. Given an arbitrary $S \in \dom(F_P)$, define $a = P(S)$. We now use the completeness of $\Oid(\apowerset(X))$ (Theorem \ref{PowerSetCompletenessTheorem}). We have $a \leq P(S)$, and $\lnot a \leq 1 = P(T)$. So there exists $S' \in \dom(\apowerset(X))$ such that $a \leq \bvs{S = S'}$ and $\lnot a \leq \bvs{T = S'}$ (we have implicitly used $\bvs{S \in \apowerset(X)} = 1$ and also for $T$ and $S'$ to simplify $\bvs{S = S'}_{\apowerset(X)}$ to $\bvs{S = S'}$ and likewise for $T$). 

We then observe that by well-definedness of $P$ with respect to $\beq$:
\[
a  = a \land a \leq P(S) \land \bvs{S = S'} \leq P(S')
\]
and also
\[
\lnot a = \lnot a \land 1 \leq \bvs{T = S'} \land P(T) \leq P(S'),
\]
so $1 = a \lor \lnot a \leq P(S')$. We now make clear that what we will use in the next part of the argument is that $P(S') = 1$ (so $P \in \dom(E_P)$) and $\bvs{S = S'} = P(S)$. 

We now prove that $\bvs{F_P \subseteq E_P} = 1$. This means we have to show that for all $S \in \dom{F_P} = \dom{\apowerset(X)}$, $F_P(S) \leq \bvs{S \in E_P}$. By definition $F_P(S) = P(S)$, while 
\[
\bvs{S \in E_P} = \bigvee_{R \in \dom{E_P}} \bvs{S = R} \land E_P(R) = \bigvee_{R \in \dom{E_P}} \bvs{S = R}
\],
as $E_P(R) = 1$. So the inequality to be proved is
\begin{equation}
\label{FinalDefeatingEmptySetEqn}
P(S) \leq \bigvee_{R \in \dom{E_P}} \bvs{S = R}.
\end{equation}
We have $S' \in \dom{E_P}$ and so
\[
P(S) = a \leq \bvs{S = S'} \leq \bigvee_{R \in \dom{E_P}} \bvs{S = S'},
\]
proving \eqref{FinalDefeatingEmptySetEqn}, and thereby showing $\bvs{E_P = F_P} = 1$. 

We can now show that $\Oid(E_P)$ is complete, using Definition \ref{CompletenessDef} (ii). We first point out that $\bvs{S_1 = S_2}_{\Oid(E_P)} = \bvs{S_1 = S_2}$ for all $S_1,S_2 \in \dom(E_P)$. Suppose there is a pairwise disjoint family of elements $(a_i)_{i \in I}$ in $A$, and a corresponding family $(S_i)_{i \in I}$ in $\dom(E_P)$ such that $a_i \leq \bvs{S_i = S_i}$ (this always holds because $\bvs{S_i = S_i} = 1$). Use completeness of $\Oid(\apowerset(X))$ (Theorem \ref{PowerSetCompletenessTheorem}) to obtain an $S \in \dom(\apowerset(X))$ such that for all $i \in I$, $a_i \leq \bvs{S_i = S}$. So for all $i \in I$
\[
a_i \land 1 \leq \bvs{S_i = S} \land P(S_i) \leq P(S).
\]
Defining $a = P(S)$, we have that there exists $S' \in \dom(E_P)$ such that $a \leq \bvs{S = S'}$, and therefore for all $i \in I$,
\[
a_i = a_i \land a \leq \bvs{S_i = S} \land \bvs{S = S'} \leq \bvs{S_i = S'}.
\]
\end{proof}

\begin{defi}
\label{BVRIDef}
If $(B,\prec)$ is an abstract base, we can represent $\RI^A(\check{B})$ as calculated in $V^A$ as follows.
\[
\RI^A(\check{B}) = \{ S : \dom(\check{B}) \rightarrow A \mid \bvs{f \text{ is a rounded ideal}} = 1 \}
\]
We find that condition (iii), being a down set, translates to the map $S$ being antitone from $\check{\prec}$ to the $\leq$ of $A$, condition (i), being non-empty, to $\bigvee_{b \in B}S(\check{b}) = 1$, and condition (ii), the other part of being directed, to 
\[
\forall x_1,x_2 \in B. S(\check{x_1}) \land S(\check{x_2}) \leq \bigvee_{x \cerp x_1,x_2} S(\check{x})
\]
For $S_1,S_2 \in \RI^A(\check{B})$, we have
\begin{align*}
\bvs{S_1 \subseteq S_2} &= \bigvee_{b \in B} S_1(\check{b}) \Rightarrow S_2(\check{b}) \\
\bvs{S_1 = S_2} &= \bigvee_{b \in B} S_1(\check{b}) \Leftrightarrow S_2(\check{b}).
\end{align*}
By Theorem \ref{SetTheoryMetaTheorem}, if we interpret the following statements in $V^A$, $\RI^A(\check{B})$ is a dcpo with base $\check{B}$ and is therefore continuous\footnote{The underlying order will not be continuous if $D$ has a bottom element, one other element, and $A(X)$ is not atomic (Proposition \ref{NotContinuousProp}).}, using $\bvs{{\subseteq}}$ as the order relation. 

There is therefore a strictly isomorphic $A$-poset that is a little more convenient to use in practice, which we write as $RI^A(B)$, which consists of functions $B \rightarrow A$ satisfying the following conditions
\begin{enumerate}[(i)]
\item $\bigvee_{b \in B} S(b) = 1$.
\item For all $b_1,b_2 \in B$, $S(b_1) \land S(b_2) \leq \bigvee_{b \cerp b_1,b_2} S(b)$. 
\item $S$ is antitone from $\prec$ to the $\leq$ of $A$, \emph{i.e.} $b_1 \prec b_2$ implies $S(b_2) \leq S(b_1)$. 
\end{enumerate}
To make $RI^A(B)$ an $A$-poset, define 
\[
\bvs{S_1 \leq S_2} = \bigwedge_{b \in B} S_1(b) \Rightarrow S_2(b).
\]
The strict isomorphism $RI^A(B) \isoarrow \RI^A(B)$ is given by composition with $\check{\blank} : B \rightarrow \check{B}$. $RI^A(B)$ and $\RI^A(B)$ are complete as setoids. 
\end{defi}
\begin{proof}
If we apply the rules for separation in $V^A$, we actually get
\[
\dom(\RI_0^A(\check{B})) = \dom(\apowerset(\check{B})) = \{ S : \dom(\check{B}) \rightarrow A \},
\]
so we use Lemma \ref{DefeatingEmptySetLemma}, taking $X = \check{B}$, and $P(S) = \bvs{S \text{ a rounded ideal}}$, and use non-emptiness of $B$ to pick a $b \in B$ and then define $T = I_{\check{b}}^A$, \emph{i.e.} for $b' \in B$
\begin{align*}
\dom(I^A_{\check{b}}) &= \dom{\check{B}} \\
I_{\check{b}}^A(\check{b'}) &= \check{\prec}(\check{b'},\check{b})^A = \iver{b' \prec b},
\end{align*}
which is a rounded ideal in the $V^A$ sense by Proposition \ref{AbstractBaseIsBaseProp}. Therefore $\bvs{\RI(\check{B}) = \RI_0(\check{B})} = 1$ and $\Oid(\RI(\check{B}))$ is complete. Transferring the definitions to the strictly isomorphic $A$-poset $RI^A(B)$ is then straightforward. 
\end{proof}

We now relate this to random variables, using some measure-theoretic results from Appendix \ref{MeasDomainAppendix}. By Lemmas \ref{GraphMbleLemma} and \ref{BorelTensorLemma}, if $D$ is a countably-based dcpo, the graph of ${\leq}$ is measurable as a subset of $D \times D$ for the product $\sigma$-algebra, which is the same as the Scott-Borel $\sigma$-algebra of the product. 

\begin{defi}
\label{RandVarDef}
Let $(X,\Sigma)$ be a measurable space and $D$ a dcpo with countable base. Define $\Ell^0(X;D)$ to be the set of $D$-valued random variables, \emph{i.e.} measurable functions $f : X \rightarrow D$, where $D$ is equipped with the Borel $\sigma$-algebra of the Scott topology (which we call the \emph{Scott-Borel} $\sigma$-algebra). 

For $a,b \in \Ell^0(X;D)$, the set $\bvs{a \leq b}_\Ell \defeq \{ x \in X \mid a(x) \leq b(x) \}$ is measurable (in $\Sigma$). It follows from antisymmetry of $\leq$ that $\{ x \in X \mid a(x) = b(x) \}$ and its complement $\{ x \in X \mid a(x) \neq b(x) \}$ are measurable. 

Now let $(X,\Sigma,\En)$ be a negligibility space. For $a,b \in \Ell^0(X;D)$, we define almost everywhere equality of random variables:
\[
a \sim b \Leftrightarrow \{ x \in X \mid a(x) \neq b(x) \} \in \En.
\]
This is an equivalence relation. We then define $L^0(X;D) = \Ell^0(X;D)/{\sim}$. 
If $a \sim a'$ and $b \sim b'$ then $\bvs{a \leq b}_\Ell \symdiff \bvs{a' \leq b}_\Ell \in \En$. We can therefore define 
\[
\bvs{[a] \leq [b]}_L = [\{ x \in X \mid a(x) \leq b(x) \}] \in A(X)
\]
\end{defi}
\begin{proof}
We first show that $\{ x \in X \mid a(x) \leq b(x) \}$ is measurable. Observe that for all $x \in X$, 
\[
x \in \langle a, b \rangle^{-1}(\graph(\leq)) \Leftrightarrow (a(x),b(x)) \in \graph(\leq) \Leftrightarrow a(x) \leq b(x),
\]
so $\{ x \in X \mid a(x) \leq b(x) \} = \langle a, b \rangle^{-1}(\graph(\leq))$. Since $\graph(\leq)$ is measurable for the product $\sigma$-algebra, $\{ x \in X \mid a(x) \leq b(x) \}$ is measurable. 

It is obvious that $\sim$ is reflexive and symmetric. For transitivity, we reason as follows, writing $\bnot S$ for $X \setminus S$ where $S \subseteq X$ throughout. Suppose $a,b,c \in L^0(X;D)$ and $a \sim b$ and $b \sim c$, so $\bnot\{ x \in X \mid a(x) = b(x) \}$ and $\bnot\{x \in X \mid b(x) = c(x) \}$ are in $\En$. Therefore, by the De Morgan dual of transitivity of equality:
\begin{align*}
\bnot \{ x \in X \mid a(x) = c(x) \} \subseteq \bnot \{ x \in X \mid a(x) = b(x) \} \cup \bnot \{ x \in X \mid b(x) = c(x) \} \in \En,
\end{align*}
so since $\En$ is an ideal, $\bnot\{ x \in X \mid a(x) = c(x) \} \in \En$, and so $a \sim c$. 

We can now show $\bvs{[a] \leq [b]}_L$ is well-defined. Suppose $a \sim a'$ and $b \sim b'$, and we consider $\{ x \in X \mid a(x) \leq b(x) \} \setminus \{x \in X \mid a'(x) \leq b'(x) \}$, aiming to prove it is in $\En$. We have
\begin{align*}
& \{ x \in X \mid a(x) \leq b(x) \} \setminus \{x \in X \mid a'(x) \leq b'(x) \} \\
={} & \{ x \in X \mid a(x) \leq b(x) \land a'(x) \not\leq b'(x) \} \\
={} & \{ x \in X \mid a(x) = a'(x) \land a(x) \leq b(x) \land a'(x) \not\leq b'(x) \} \\
 &\cup \{ x \in X \mid a(x) \neq a'(x) \land a(x) \leq b(x) \land a'(x) \not\leq b'(x) \} \\
={} & \{ x \in X \mid b(x) = b'(x) \land a(x) = a'(x) \land a(x) \leq b(x) \land a'(x) \not\leq b'(x) \} \\
 &\cup \{ x \in X \mid b(x) \neq b'(x) \land a(x) = a'(x) \land a(x) \leq b(x) \land a'(x) \not\leq b'(x) \} \\
 &\cup \{ x \in X \mid a(x) \neq a'(x) \land a(x) \leq b(x) \land a'(x) \not\leq b'(x) \}
\end{align*}
The bottom two sets in the ternary union are subsets of elements of $\En$, so are in $\En$. The top set is equal to
\[
\{ x \in X \mid b(x) = b'(x) \land a(x) = a'(x) \land a(x) \leq b(x) \land a(x) \not\leq b(x) \} = \emptyset \in \En.
\]
The proof for the set difference in the other direction is similar, and since $\En$ is closed under countable unions, this shows that the symmetric difference is in $\En$, as required. 
\end{proof}

The map $F_{X,B}$ serves the purpose that $G_X$ \eqref{GXDefEqn} served for $\powerset(Y)$-valued random variables, though it doesn't quite specialize to it in the case that $B = \pfin(Y)$ for $Y$ a countable set. 

\begin{defi}
Let $(X,\Sigma,\En)$ be a negligibility space, and $D$ be a continuous dcpo with a countable base $B$. We define a map $F_{X,B} : L^0(X;D) \rightarrow RI^{A(X)}(B)$ as follows. Given $a \in \Ell^0(X;D)$
\[
F_{X,B}([a]) = [a^{-1}(\uu b)]
\]
and this is well-defined with respect to equality almost everywhere. 
\end{defi}
\begin{proof}
We first show that if $a \sim a'$, then $a^{-1}(\uu b) \triangle a'^{-1}(\uu b) \in \En$. Define $N = \{ x \in X \mid a(x) \neq a'(x) \}$, and $N \in \En$ by the definition of $\sim$. So $a|_{X \setminus N} = a'|_{X \setminus N}$, and therefore 
\[
a^{-1}(\uu b) \cap (X \setminus N) = a|_{X \setminus N}^{-1}(\uu b) = a'|_{X \setminus N}^{-1}(\uu b) = a'^{-1}(\uu b) \cap X \setminus N
\]
Then
\begin{align*}
a^{-1}(\uu b) \setminus a'^{-1}(\uu b) &= ((a^{-1}(\uu b) \cap (X \setminus N)) \cup (a^{-1}(\uu b) \cap N)) \setminus a'^{-1}(\uu b) \\
 &= ((a^{-1}(\uu b) \cap (X \setminus N)) \setminus a'^{-1}(\uu b)) \cup ((a^{-1}(\uu b) \cap N) \setminus a'^{-1}(\uu b)) \\
 &= ((a'^{-1}(\uu b) \cap (X \setminus N)) \setminus a'^{-1}(\uu b)) \cup ((a^{-1}(\uu b) \cap N) \setminus a'^{-1}(\uu b)) \\
 &= \emptyset \cup ((a^{-1}(\uu b) \cap N) \setminus a'^{-1}(\uu b)) \\
 &= a^{-1}(\uu b) \cap N \cap (X \setminus a'^{-1}(\uu b)) \subseteq N,
\end{align*}
so $a^{-1}(\uu b) \setminus a'^{-1}(\uu b) \in \En$. The argument for the other set difference is similar, so $a^{-1}(\uu b) \symdiff a'^{-1}(\uu b) \in \En$. 

We now have to show that $F_{X,D}([a]) \in RI^A(B)$, so we use (i)-(iii) from Definition \ref{BVRIDef}. 
\begin{enumerate}[(i)]
\item \begin{align*}
\bigvee_{b \in B} F_{X,D}([a])(b) &= \bigvee_{b \in B} [a^{-1}(\uu b)] \\
 &= \left[\bigcup_{b \in B} a^{-1}(\uu b)\right] & \text{since } B \text{ countable} \\
 &= \left[a^{-1}\left(\bigcup_{b \in B} \uu b\right)\right] \\
 &= [a^{-1}(D)] = [X] = 1,
\end{align*}
because $\bigcup_{b \in B} \uu b = D$ as $b$ is a base, so each element $d \in D$ has an element $b \in B$ such that $b \ll d$. 
\item Let $b_1,b_2 \in B$. We have
\begin{align*}
F_{X,D}([a])(b_1) \land F_{X,D}([a])(b_2) &= [a^{-1}(\uu b_1)] \land [a^{-1}(\uu b_2)] \\
 &= [a^{-1}(\uu b_1 \cap \uu b_2)] \\
 &= \left[a^{-1}\left(\bigcup_{b \gg b_1,b_2} \uu b\right)\right] & \text{see below for explanation} \\
 &= \left[\bigcup_{b \gg b_1,b_2} a^{-1}(\uu b)\right] \\
 &= \bigvee_{b \gg b_1,b_2} [a^{-1}(\uu b)] & B \text{ countable} \\
 &= \bigvee_{b \gg b_1,b_2} F_{X,D}([a])(b),
\end{align*}
where $\uu b \cap \uu b_2 = \bigcup_{b \gg b_1,b_2} \uu b$ by part (iv) of Definition \ref{AbstractBaseDef}, by Proposition \ref{AbstractBaseIsBaseProp}. 
\item Let $b_1,b_2 \in B$, such that $b_1 \ll b_2$. By transitivity of $\ll$, $\uu b_2 \subseteq \uu b_1,$ so 
\[
F_{X,B}([a])(b_2) = [a^{-1}(\uu b_2)] \leq [a^{-1}(\uu b_1)] = F_{X,B}([a])(b_1).
\]
\end{enumerate}
\end{proof}

This definition makes $(L^0(X;D),\bleq)$ into an $A(X)$-valued poset, \emph{i.e.} it satisfies the two axioms in Definition \ref{BVPosetDef}. 

\begin{thm}
\label{RandVarsContTheorem}
Let $(X,\Sigma,\En)$ be a negligibility space, $D$ a continuous dcpo with a countable base $B$. $(L^0(X;D),\bleq_L)$ is an $A$-poset whose $A$-setoid is complete, for each  the map $F_{X,B}$ is a strict poset isomorphism $L^0(X;D) \isoarrow RI^A(B)$. Since the structure of $(L^0(X;D),\bleq_L)$ does not depend on $B$, this proves that if $B'$ is a countable base for for $D$, we have $RI^A(B) \cong RI^A(B')$ as a strict isomorphism. Since $RI^A(B) \cong \RI^A(\check{B})$ by a strict isomorphism, we can take any statement provable of dcpos with countable bases in $V$ and have it in the $A$-valued sense for $L^0(X;D)$. 
\end{thm}
\begin{proof}
By Lemma \ref{BuildPosetByIsoLemma}, we only need to show that $F_{X,B}$ is a bijection such that for all $[a],[b] \in L^0(X;D)$
\begin{equation}
\label{LPosetEmbedEqn}
\bvs{[a] \leq [b]}_L = \bvs{F_{X,B}([a]) \leq F_{X,B}([b])}_{RI^A(B)}.
\end{equation}
In a domain $D$ with a base $B$ with $x_1,x_2 \in D$, $x_1 \leq x_2$ iff for all $b \in B$, $b \ll x_1$ implies $b \ll x_2$ Lemma \ref{BaseIneqLemma}. As in the proof of Lemma \ref{GraphMbleLemma}
\[
\{ (d_1,d_2) \in D\times D \mid d_1 \leq d_2 \} = \bigcap_{b \in B} ((D \setminus \uu b) \times D) \cup (D \times (D \setminus \uu b)).
\]
So
\begin{align*}
\bvs{[a_1] \leq [a_2]}_L &= [\{ x \in X \mid a_1(x) \leq a_2(x) \}] \\
 &= [\langle a_1, a_2 \rangle^{-1}(\graph(\leq))] \\
 &= \left[\langle a_1, a_2 \rangle^{-1} \left(\bigcap_{b \in B} ((D \setminus \uu b) \times D) \cup (D \times \uu b)\right)\right] \\
 &= \bigwedge_{b \in B} [\langle a_1, a_2 \rangle^{-1}((D \setminus \uu b) \times D)] \lor [\langle a_1, a_2 \rangle^{-1}(D \times \uu b)] \\
 &= \bigwedge_{b \in B} [a_1^{-1}(D \setminus \uu b)] \lor [a_2^{-1}(\uu b)] \\
 &= \bigwedge_{b \in B} [a_1^{-1}(\uu b)] \Rightarrow [a_2^{-1}(\uu b)] \\
 &= \bigwedge_{n \in B} F_{X,B}(a_1)(b) \Rightarrow F_{X,B}(a_2)(b) \\
 &= \bvs{F_{X,B}(a_1) \leq F_{X,B}(a_2)}_{RI^A(B)}.
\end{align*}
where we have used the countability of $B$ to turn the $\bigcap$ to a $\bigwedge$. 

We now observe that $L(X;B)$ is strict, as follows. First see that
\begin{align*}
\bvs{[a] = [b]}_L &= \bvs{[a] \leq [b]}_L \land \bvs{[b] \leq [a]}_L \\
 &= [\{ x \in X \mid a(x) \leq b(x) \}] \land [\{ x \in X \mid b(x) \leq a(x) \}] \\
 &= [\{ x \in X \mid a(x) = b(x) \}],
\end{align*}
so if $\bvs{[a] = [b]}_L = 1$, then $\{ x \in X \mid a(x) \neq b(x) \} \in \En$, so $[a] = [b]$. 

Since $\bvs{[a] = [b]}_L = \bvs{F_{X,B}([a]) = F_{X,B}([b])}$, and both are strict $A$-posets, it follows that $F_{X,B}$ is injective as a function. 

We prove surjectivity of $F_{X,B}$ as follows. Suppose $S : B \rightarrow A(X)$ is in $RI^A(B)$. We start by building a map $S' : B \rightarrow \Sigma$ such that for all $b_1,b_2 \in B$ we have
\begin{equation}
\label{SPrimeEqn}
S'(b_1) \cap S'(b_2) = \bigcup_{b \gg b_1,b_2} S'(b),
\end{equation}
as actual equality, instead of just in $A(X)$. So for each $b \in B$, pick an $S_b \in \Sigma$ such that $[S_b] = S(b)$. Now, parts (ii) and (iii) together imply that for all $b_1,b_2 \in B$, $S(b_1) \land S(b_2) = \bigvee_{b \gg b_1,b_2} S(b)$, so the set $N_{b_1,b_2} \defeq S_{b_1} \cap S_{b_2} \symdiff \bigcup_{b \gg b_1,b_2} S_b$ belongs to $\En$. Define $N = \bigcup_{b_1,b_2 \in B} N_{b_1,b_2}$, and as a countable union $N \in \En$. Define $S'(b) = S_b \cup N$, and we have $[S'(b)] = [S_b] = S(b)$. 

We aim to show that for all $b_1,b_2 \in B$, \eqref{SPrimeEqn} holds. We do this by showing the set difference each way is $\emptyset$. We use $\bnot S$ for $X \setminus S$ to simplify notation. 
\begin{align*}
& S'(b_1) \cap S'(b_2) \setminus \bigcup_{b \gg b_1,b_2} S'(b) \\
 &= (S_{b_1} \cup N) \cap (S_{b_2} \cup N) \cap \bnot \left(\bigcup_{b \gg b_1,b_2} S_b \cup N\right) \\
 &= ((S_{b_1} \cap S_{b_2}) \cup N) \cap \left(\bnot\bigcup_{b \gg b_1,b_2}S_b\right) \cap \bnot N \\
 &= (S_{b_1} \cap S_{b_2}) \cap \left(\bnot\bigcup_{b \gg b_1,b_2} S_b\right) \cap \bnot N \\
 &= \left((S_{b_1} \cap S_{b_2}) \setminus \bigcup_{b \gg b_1,b_2} S_b\right) \cap \bnot N \\
 &\subseteq \left((S_{b_1} \cap S_{b_2}) \setminus \bigcup_{b \gg b_1,b_2} S_b\right) \cap \bnot N_{b_1,b_2} \\
 &= \emptyset,
\end{align*}
and the other way
\begin{align*}
& \left(\bigcup_{b \gg b_1,b_2}S'(b)\right)\setminus (S'(b_1) \cap S'(b_2)) \\
 &= \left(\bigcup_{b \gg b_1,b_2} S_b \cup N \right) \setminus ((S'(b_1) \cup N) \cap (S'(b_2) \cup N)) \\
 &= \left(\bigcup_{b \gg b_1,b_2} S_b \cup N \right) \cap \bnot (S_{b_1} \cap S_{b_2}) \cup N) \\
 &= \left(\left(\bigcup_{b \gg b_1,b_2} S_b\right) \cup N\right) \cap \bnot (S_{b_1} \cap S_{b_2}) \cap \bnot N \\
 &= \left(\left(\bigcup_{b \gg b_1,b_2} S_b\right) \setminus (S_{b_1} \cap S_{b_2})\right) \cap \bnot N \\
 &= \left(\left(\bigcup_{b \gg b_1,b_2} S_b\right) \setminus (S_{b_1} \cap S_{b_2})\right) \cap \bnot N_{b_1,b_2} \\
 &= \emptyset.
\end{align*}

By Theorem \ref{ScottBorelPresentationTheorem}, $S'$ extends to a $\sigma$-Boolean homomorphism $\Bo(D) \rightarrow \Sigma$. It then follows by Sikorski duality \cite[Proposition 9 (ii)]{unresStoneProc} that there exists a measurable function $f : (X,\Sigma) \rightarrow (D,\Bo(D))$ such that for all $b \in B$, $f^{-1}(\uu b) = S'(b)$, so 
\[
F_{X,B}([f]) = [f^{-1}(\uu b)] = [S'(b)] = [S_b \cup N] = [S_b] = S(b),
\]
finishing the proof of surjectivity. 
\end{proof}

It follows that we could use random variables taking values in \emph{e.g.} Scott's continuous lattice $D_\infty$ (see \emph{e.g.} \cite[\S 18.2]{barendregt84}) as the reflexive continuous lattice with numerals in Theorem \ref{IncomparableDegreesTheorem}, and more generally, we can extend the methods described above to countably-based domains. 

\section{Conclusions}
We have developed domain theory in a Boolean-valued universe of sets.  Using the
measure algebra as the Boolean algebra we obtained a domain of random variables
which can be seen to be a reflexive domain \emph{in the internal language of the
	Boolean-valued set theory}.  We have focused on the pure $\lambda$-calculus
here but it should be straightforward to extend this to a $\lambda$-calculus
with probabilistic choice as an explicit primitive.  

There are a number of directions for future work.  First, one can use this construction to give semantics to a $\lambda$-calculus extended with probabilistic choice as was done in \cite{Bacci18b}.  In this model it will be interesting to see which equations involving the interplay of choice and the standard $\lambda$-calculus constructions are valid.  One could then relate it to an operational semantics as, for example, in \cite{Amorim21} but one would need a Boolean-valued notion of operational semantics.  More interestingly one could define conditioning as a primitive and explore its semantics.
A second line of research is the notion of approximation.  A notion of ``approximate equality'' has been developed recently \cite{Mardare16}; the connection to the notion of equality used in the present paper is unclear but there is a similarity in that in both cases equality may only hold partially.


\bibliographystyle{alphaurl}
\bibliography{dvrvconf}

\appendix

\section{Measure-Theoretic Results in Domain Theory}
\label{MeasDomainAppendix}

The purpose of this appendix is to collect some results about the Borel $\sigma$-algebra of a countably-based domain. The following lemma is well-known, but included for completeness. 
\begin{lem}
\label{BaseIneqLemma}
Let $D$ be a poset with base $B$. For all $d, d' \in D$, $d \leq d'$ iff for all $b \in B$, $b \ll d$ implies $b \ll d'$. 
\end{lem}
\begin{proof}
Clearly, if $d \leq d'$, then $b \ll d$ implies $b \ll d'$ by the compatibility of $\ll$ with $\leq$. For the other direction, if $b \ll d$ implies $b \ll d'$, then $\dd d \cap B \subseteq \dd d' \cap B$, and therefore $d'$ is an upper bound for $\dd d \cap B$ and $d' \geq d$. 
\end{proof}

We require a description of countably presented $\sigma$-Boolean algebras. As in \cite[\S IV]{unresStoneProc}, given a set $X$ we use $H(X) = \Ba(2^X)$ to refer to the free $\sigma$-Boolean algebra on $X$, which is the Baire $\sigma$-algebra of the compact Hausdorff space $2^X$. In the case that $X$ is countable, $2^X$ is metrizable, so $\Ba(2^X) = \Bo(2^X)$. We use $\eta_X$ for the map $X \rightarrow \Ba(2^X)$ defined by $\eta_X(x) = \{ a : X \rightarrow 2 \mid a(x) = 1 \}$. In the case of a function $f : X \rightarrow B$ where $X$ is a set and $B$ a $\sigma$-Boolean algebra, we use $\tilde{f} : H(X) \rightarrow B$ for the unique extension along $\eta_X$. 

We also use the categorical version of Sikorski's duality between $\sigma$-Boolean algebras and measurable spaces described in \cite[\S III]{unresStoneProc}, changing the notation slightly. Given a measurable space $(X,\Sigma)$, we define $\sigma(X,\Sigma) = \Sigma$, which is a $\sigma$-Boolean algebra, and given a measurable function $f : (X,\Sigma) \rightarrow (Y,\Theta)$ define\footnote{In \cite[\S III]{unresStoneProc} this was called $F$.} $\sigma(f)(T) = f^{-1}(T)$, which defines a $\sigma$-Boolean algebra homomorphism $\Theta \rightarrow \Sigma$. Given a $\sigma$-Boolean algebra $A$, we define the underlying set of its \emph{Sikorski space} $\Sik(A)$ to be the set of $\sigma$-ultrafilters on $A$, \emph{i.e.} ultrafilters closed under countable meets. For $a \in A$, define $\ban{a}$ to be the set of $\sigma$-ultrafilters $u$ such that $a \in u$. We define the $\sigma$-field of the Sikorski space to be the image of $\ban{\blank}$, so $\Sik(A)$ is a measurable space. If $f : A \rightarrow B$ is a $\sigma$-Boolean homomorphism and $v \in \Sik(B)$, we define\footnote{In \cite[\S III]{unresStoneProc}, this functor is called $G$.} $\Sik(f)(v) = f^{-1}(v)$, and this defines a measurable function $\Sik(f) : \Sik(B) \rightarrow \Sik(A)$. In \cite[\S III, Theorem 5]{unresStoneProc} it is proved that $\sigma : \Mble \rightarrow \sigmaBA\op$ and $\Sik : \sigmaBA\op \rightarrow \Mble$ are functors and $\sigma$ is a left adjoint to $\Sik$. The counit is $\ban{\blank}$ and the unit is $\princip{\blank}$, where, for a measurable space $(X,\Sigma)$ and $x \in X$, $\princip{x} = \{S \in \Sigma \mid x \in S\}$. 

 We say that a measurable space $(X,\Sigma)$ is \emph{separated} if for each pair of distinct points $x,y \in X$ there exists $S \in \Sigma$ such that $x \in S$ and $y \not\in S$. Because $\sigma$-fields are closed under complement, this is symmetric in $x$ and $y$. Clearly $(X,\Sigma)$ is separated iff $\princip{\blank}$ is injective. If $\princip{\blank}$ is also surjective, \emph{i.e.} for each $\sigma$-ultrafilter $u$ on $\Sigma$ there is an $x \in X$ such that $u = \princip{x}$, we say, following Sikorski, that $(X,\Sigma)$ is $\sigma$-perfect, and in this case it is also true that $\princip{\blank} : (X,\Sigma) \rightarrow \Sik(\Sigma)$ is a measurable isomorphism\footnote{It is not the case that a bijective measurable function is a measurable isomorphism in general.}. The full subcategory of $\Mble$ on $\sigma$-perfect measurable spaces will be called $\PMble$.

If a $\sigma$-Boolean algebra $A$ is separated by its $\sigma$-ultrafilters, \emph{i.e.} if every non-zero element is contained in a $\sigma$-ultrafilter, then $\ban{\blank}$ is an isomorphism $A \rightarrow \sigma(\Sik(A))$. We call such $\sigma$-Boolean algebras $\sigma$-spatial, by analogy to the theory of locales, and they form a full subcategory $\sigmaBASp$ of $\sigmaBA$. We have that $\sigma$ and $\Sik$ define an adjoint equivalence $\PMble \simeq \sigmaBASp$. 

In the following, if $A$ is a $\sigma$-Boolean algebra, and $X \subseteq A$, we write $I(X)$ for the $\sigma$-ideal generated by $X$, \emph{i.e.} the set 
\[
I(X) = \left\{a \in A \,\left|\, \exists (x_i)_{i \in I} . x_i \in X \text{ and } a \leq \bigvee_{i \in I}x_i \right. \right\},
\]
and we use $[\blank]_X : A \rightarrow A/I(X)$ for the quotient mapping. If $X$ is countable, then $I(X) = I\left(\left\{\bigvee X\right\}\right)$, \emph{i.e.} countably generated $\sigma$-ideals are principal.

A \emph{presentation} of a $\sigma$-Boolean algebra $A$ is a triple $(G,f,R)$ where $G$ is a set, $f: G \rightarrow A$ is a function, and $R \subseteq H(G)$ such that $f(G)$ generates $A$, \emph{i.e.} $\tilde{f}(G) = A$, and such that $I(R) = \tilde{f}^{-1}(\bot)$. We say a presentation $(G,f,R)$ is \emph{countable} if $G$ and $R$ are countable. Given a pair $(G,R)$ where $G$ is a set and $R \subseteq H(G)$, $(G,\eta_G,R)$ is a presentation of $H(G)/I(R)$, so we will also call $(G,R)$ is a presentation, the algebra $H(G)/I(R)$ being implicit. If $(G,f,R)$ is a presentation of $A$, then there is a unique isomorphism $i : (H(G)/I(R), [\blank]_R) \rightarrow (A,f)$ in the comma category $H(G)\downarrow \sigmaBA$, so this does not cause any problems.

We say that $\Sigma$ is \emph{countably generated} if there exists a set $(S_i)_{i \in I}$ with $I$ countable that generates $\Sigma$ as a $\sigma$-field. A family of measurable sets $(T_i)_{i \in I}$ is said to be a \emph{separating family} if the statement that $x \in T_i \Leftrightarrow y \in T_i$ holds for all $i \in I$ implies that $x = y$. A measurable space is \emph{countably separated} if it has a countable separating family. We will be using the following result without comment in future, but we give the proof now for completeness.
\begin{lem}
\label{GenFamImpliesSepFamLemma}
Let $(X,\Sigma)$ be a separated $\sigma$-field generated by $(G_i)_{i \in I}$. Then $(G_i)_{i \in I}$ is a separating family for $(X,\Sigma)$. In particular, a separated countably generated space is countably separated. 
\end{lem}
\begin{proof}
Let $x,y \in X$ and suppose that for all $i \in I$, $x \in G_i \Leftrightarrow y \in G_i$. We need to show that $x = y$. First we show that for all $S \in \Sigma$, $x \in S \Leftrightarrow y \in S$. Let $\Sigma' = \{ S \in \Sigma \mid x \in S \Leftrightarrow y \in S \}$. By our initial assumption, for all $i \in I$, $G_i \in \Sigma'$. If $S \in \Sigma'$, then $x \in S \Leftrightarrow y \in S$, so $x \not\in S \Leftrightarrow y \not\in S$, so $x \in X \setminus S \Leftrightarrow y \in X \setminus S$. This proves that $\Sigma'$ is closed under complement.

Let $(S_j)_{j \in \N}$ be a sequence of elements of $\Sigma'$, which is to say, $x \in S_j \Leftrightarrow y \in S_j$ for all $j \in \N$. Then
\[
x \in \bigcup_{j = 1}^\infty S_j \Leftrightarrow \exists j \in \N . x \in S_j \Leftrightarrow \exists j \in \N . y \in S_j \Leftrightarrow y \in \bigcup_{i=1}^\infty S_j,
\]
so $\bigcup_{j = 1}^\infty S_j \in \Sigma'$. Therefore $\Sigma'$ is a $\sigma$-subfield of $\Sigma$ containing $G_i$ for all $i \in I$, so $\Sigma' = \Sigma$. Therefore, for all $S \in \Sigma$, $x \in S \Leftrightarrow y \in S$, which, as $(X,\Sigma)$ is separated, implies $x =y$. 
\end{proof}

We use the following lemma to prove that certain $\sigma$-homomorphisms between $\sigma$-fields arise from embeddings of measure spaces.
\begin{lem}
\label{PrincipalMbleEmbeddingLemma}
Let $(Y,\Theta)$ be a $\sigma$-perfect measurable space, and $(X,\Sigma)$ a separated measurable space, and $R \in \Theta$. Suppose that $f : \Theta \rightarrow \Sigma$ is a $\sigma$-homomorphism that is surjective and such that $f(R) = \emptyset$, and let $g : (X,\Sigma) \rightarrow (Y,\Theta)$ be the unique measurable map such that $\sigma(g) = f$. Then if for all $y \in Y \setminus R$ there exists an $x \in X$ such that $g(x) = y$, $I(R) = f^{-1}(\emptyset)$ and $g$ is an embedding of $(X,\Sigma)$ into $(Y,\Theta)$ with image $Y \setminus R$, \emph{i.e.} it restricts to an isomorphism $(X,\Sigma) \rightarrow (Y \setminus R, \Theta|_{Y\setminus R})$. 
\end{lem}
\begin{proof}
We first remark that $g$ can be defined to be $\princip{\blank}^{-1}_Y \circ \Sik(f) \circ \princip{\blank}_X$, which only requires $(Y,\Theta)$ to be $\sigma$-perfect. It is unique because if $g,h : (X,\Sigma) \rightarrow (Y,\Theta)$ satisfy $\sigma(g) = \sigma(h)$ for all $T \in \Theta$ and $x \in X$ we have $x \in g^{-1}(T) \Leftrightarrow x \in h^{-1}(T)$, so $g(x) \in T \Leftrightarrow h(x) \in T$. As $\sigma$-perfect spaces are separated, $g(x) = h(x)$, so $g = h$. 

As $f(R) = \emptyset$, we have $I(R) \subseteq f^{-1}(\emptyset)$, so to show that $I(R) = f^{-1}(\emptyset)$, it suffices to show that $f(T) = \emptyset$ implies $T \subseteq R$, for any $T \in \Theta$. We show the contrapositive, that $T \not\subseteq R$ implies $f(T) \neq \emptyset$. 

Let $y \in T \setminus R$. Since $y \in Y \setminus R$, there exists $x \in X$ such that $g(x) = y$. So $x \in g^{-1}(y) \subseteq g^{-1}(T) = f(T)$, proving $f(T) \neq \emptyset$. 

Now we need to show that $g$ is a measurable embedding with image $Y \setminus R$. By definition it is measurable. The hypothesis on $g$ from the statement of the lemma implies directly that $Y \setminus R \subseteq g(X)$. If there were $x \in X$ such that $g(x) \in R$, then $f(R) = g^{-1}(R) \ni x$, contradicting $f(R) = \emptyset$, so $g(X) = Y \setminus R$. 

We show that $g$ is injective as follows. Suppose that there exist $x,x' \in X$ such that $x \neq x'$. As $(X,\Sigma)$ is separated, there exists $S \in \Sigma$ such that $x \in S$ and $x' \not\in S$. As $f$ is surjective, there exists $T \in \Theta$ such that $f(T) = S$, \emph{i.e.} $g^{-1}(T) = S$. Therefore $g(x) \in T$. If $g(x') \in T$, then $x' \in g^{-1}(S)$, which is a contradiction, so $g(x') \not\in T$, and therefore $g(x) \neq g(x')$. 

We show that $g$ is a measurable embedding, or equivalently an isomorphism onto its image, as follows. Let $S \in \Sigma$. As $f$ is surjective, there exists $T \in \Theta$ such that $S = f(T) = g^{-1}(T)$. As $g(X) = Y \setminus R$, $S = g^{-1}(T \cap (Y \setminus R))$. As $g$ is a bijection between $X$ and $Y \setminus R$, $g(S) = T \cap (Y \setminus R)$. This implies that the set-theoretic inverse of $g$ from $Y \setminus R \rightarrow X$ is measurable. 
\end{proof}

The following is useful for proving that certain triples $(G,f,R)$ are actually presentations, and is an extension of the result \cite[Lemma 10]{unresStoneProc}. It has the side effect of proving that the space being presented is a standard Borel space.
\begin{prop}
\label{CountablePresentationImpliesStandardProp}
Let $(X,\Sigma)$ be a separated measure space, $G$ a countable set, $f : G \rightarrow \Sigma$ a function and $R \subseteq H(G)$ a countable set such that:
\begin{enumerate}[(i)]
\item $f(G)$ generates $\Sigma$.
\item For all $r \in R$, $\tilde{f}(r) = \emptyset$
\item For all $s \in 2^G$ such that for all $r \in R$, $s \not\in r$, there exists $x \in X$ such that for all $g \in G$, $x \in f(g) \Leftrightarrow s(g) = 1$.
\end{enumerate}
Then $(G,f,R)$ is a countable presentation of $(X,\Sigma)$, and the unique map $h : (X,\Sigma) \rightarrow (2^G,\Bo(2^G))$ such that $\sigma(h) = \tilde{f}$ is a measurable embedding with image $2^G \setminus \bigcup_{r \in R} r$, proving $(X,\Sigma)$ is a standard Borel space. 
\end{prop}
\begin{proof}
By (i) and (ii), we have that $f(G)$ generates $\Sigma$ (\emph{i.e.} $\tilde{f}$ is surjective). We define $\overline{r} = \bigvee R$, which is an element of $H(G)$ because $R$ is countable, and observe that $I(R) = I(\overline{r})$, and $\tilde{f}(\overline{r}) = \emptyset$, by (ii). By (iii), we have that for all $s \in 2^G \setminus \overline{r}$, there exists $x \in X$ such that $x \in f(g) \Leftrightarrow s(g) = 1$. Let $h : X \rightarrow 2^G$ be the unique measurable map such that $\sigma(h) = \tilde{f}$. We have that
\begin{align*}
x \in f(g) & \Leftrightarrow x \in \tilde{f}(\eta_G(g)) & \text{definition of } \tilde{f} \\
 &\Leftrightarrow x \in \sigma(h)(\eta_G(g)) & \text{definition of } h \\
 &\Leftrightarrow x \in h^{-1}(\eta_G(g)) \\
 &\Leftrightarrow h(x) \in \eta_G(g) \\
 &\Leftrightarrow h(x)(g) = 1 &\text{definition of } \eta_G(g),
\end{align*}
so $h(x)(g) = 1 \Leftrightarrow s(g) = 1$. As $s$ takes values in $2$, this implies $h(x) = s$. 

Therefore we can apply Lemma \ref{PrincipalMbleEmbeddingLemma} to conclude that $I(\overline{r}) = \tilde{f}^{-1}(\emptyset)$, so $(G,f,R)$ is a presentation of $\Sigma$, and $h$ embeds $(X,\Sigma)$ onto $(2^G \setminus \overline{r}, H(G)|_{2^G \setminus \overline{r}})$. It follows that $(X,\Sigma)$ is a standard Borel space because $(2^G,H(G))$ is a standard Borel space ($G$ is countable), and any Borel subspace of a standard Borel space is a standard Borel space.
\end{proof}

\begin{lem}
\label{UpsetBorelLemma}
Let $D$ be a continuous dcpo with countable basis. Then for all $d \in D$, the set $\upset d$ is in $\Bo(D)$, where $D$ has the Scott topology. 
\end{lem}
\begin{proof}
Let $B$ be a countable basis for $D$. Then
\begin{align*}
\upset d &= \{ d' \in D \mid d \leq d' \} \\
 &= \{ d' \in D \mid \forall b \in B. b \ll d \Rightarrow b \ll d' \} & \text{Lemma \ref{BaseIneqLemma}} \\
 &= \{ d' \in D \mid \forall b \in \dd d \cap B . d' \in \uu b \} \\
 &= \bigcap_{b \in \dd d \cap B} \uu b.
\end{align*}
As $B$ is countable, this expresses $\upset b$ as the countable intersection of Scott open sets, which is therefore a Borel set. 
\end{proof}

For ease of notation, we will write $\Sigma_D$ for $\Bo(D)$. 

If $D$ is a continuous domain with countable base $B$, we define $f : B \rightarrow \Sigma_D$ by $f(b) = \uu b$. This extends to a $\sigma$-homomorphism $\tilde{f} : H(B) \rightarrow \Sigma_D$. To ease notation, we write $g_b$ for $\eta_B(b)$. We can define 
\[
R_B = \left\{ \left. (g_b \cap g_{b'}) \triangle \bigcup_{b'' \gg b,b'} g_{b''} \,\right| \, b,b' \in B \right\}.
\]
This is a countable subset of $H(B)$. 

\begin{thm}
\label{ScottBorelPresentationTheorem}
$(B,f,R_B)$ is a countable presentation of $\Sigma_D$, so $(D,\Sigma_D)$ is a standard Borel space. 
\end{thm}
\begin{proof}
We use Proposition \ref{CountablePresentationImpliesStandardProp}. We first show that $(D,\Sigma_D)$ is separated. Let $d,d' \in D$ be distinct elements. As $d = \bigvee \dd d \cap B$ and $d' = \bigvee \dd d' \cap B$, there must be some $b \in B$ such that $b \ll d$ and $b \not\ll d'$, or the other way round. Therefore $d \in \uu b$ and $d' \not\in \uu b$, or vice-versa, so $\Sigma_D$ is separated. 

Since $\{ \uu b \mid b \in B \}$ is a countable base for the Scott topology, every Scott-open set in $D$ is a countable union of elements of the form $\uu b$, where $b \in B$, so $f(B)$ generates $\Sigma_D$ as a $\sigma$-field, proving (i). The elements of $R_B$ are of the form $g_b \cap g_{b'} \triangle \bigcup_{b'' \gg b,b'}g_{b''}$ where $b,b' \in B$. To prove (ii), it suffices to show that $\uu b \cap \uu b' = \bigcup_{b'' \gg b,b'}\uu b''$. If $b'' \gg b,b'$, and $d \gg b''$, then $d \gg b,b'$. Therefore $\bigcup_{b'' \gg b,b'}\uu b'' \subseteq \uu b \cap \uu b'$. For the opposite inclusion, suppose that $d \in \uu b \cap \uu b'$, \emph{i.e.} $d \gg b,b'$. Since $d = \bigvee \dd d \cap B$ and $\uu b \cap \uu b'$ is Scott open, there exists $b'' \in \dd d \cap B$ such that $b \in \uu b \cap \uu b'$. It follows that $b'' \in B$, and $b,b' \ll b'' \ll d$, \emph{i.e.} $d \in \uu b''$, and therefore $d \in \bigcup_{b'' \gg b,b}\uu b''$. This proves (ii).

All that remains is to prove (iii). Let $s \in 2^B$ such that $s \not\in r$ for all $r \in R_B$. We need to produce an element $d \in D$ such that for all $b \in B$, $d \in \uu b \Leftrightarrow s(b) = 1$. To this end, we want to define
\[
d = \bigvee \{ b \in B \mid s(b) = 1 \},
\]
which will be a definition if $\{ b \in B \mid s(b) = 1 \}$ is directed, so we prove this. If $b,b' \in B$ such that $s(b) = 1 = s(b')$, then $s \in g_b \cap g_{b'}$. Since $s \not\in g_b \cap g_{b'} \triangle \bigcup_{b'' \gg b,b'} g_{b''}$, $s \in \bigcup_{b'' \gg b,b'}g_{b''}$, \emph{i.e.} there exists $b'' \in B$ such that $b'' \gg b,b'$ and $s(b'') = 1$. Importantly for later, this shows that $\{b \in B \mid s(b) = 1 \}$ is not just $\leq$-directed, but $\ll$-directed. 

Therefore $d$ is defined as the supremum of a directed set in $D$. We need to show that $d \in \uu b \Leftrightarrow s(b) = 1$. If $d \in \uu b$, then $b \leq d$, so $s(b) = 1$ by the definition of $d$. For the other direction, if $s(b) = 1$, then as $\{ b \in B \mid s(b) = 1\}$ is $\dd$-directed, there exists $b' \in B$ such that $b \ll b'$ and $s(b') = 1$. Therefore $b' \leq d$ and $b \ll b'$, so $d \in \uu b$. 
\end{proof}

By \cite[Proposition 9 (ii)]{unresStoneProc}, if $(X,\Sigma)$ is a $\sigma$-perfect measurable space, measurable functions $f : (X,\Sigma) \rightarrow (D,\Sigma_D)$ are in bijection with functions $g : B \rightarrow \Sigma$ such that for each pair $b,b' \in B$, $\bigvee_{b'' \gg b,b'}g(b'') = g(b) \land g(b')$, where for a measurable function $f$ the corresponding $g$ is defined by $g(b) = f^{-1}(\uu b)$.

We finish with some other facts that are needed in the main text. 

\begin{lem}
\label{GraphMbleLemma}
Let $D$ be a continuous dcpo with a countable basis. Then the graph of $\leq$,
\[
\graph(\leq) = \{ (d,e) \in D \times D \mid d \leq e \}
\]
is measurable in $(D \times D, \Sigma_D \otimes \Sigma_D)$. It follows that the analogous sets with $d \not\leq e$, $d = e$ and $d \neq e$ are also measurable.
\end{lem}
\begin{proof}
Recall that $\Sigma_D \otimes \Sigma_D$ is the $\sigma$-field generated by rectangles, \emph{i.e.} sets of the form $S \times T$ for $S,T \in \Sigma_D$. If we take $B$ to be a countable basis for $D$, we have
\begin{align*}
\{ (d,e) \in D \times D \mid d \leq e \} &= \{ (d,e) \in D \times D \mid \forall b \in B. b \ll d \Rightarrow b \ll e \} & \text{Lemma \ref{BaseIneqLemma}} \\
 &= \bigcap_{b \in B} \{ (d,e) \in D \times D \mid d \in \uu b \Rightarrow e \in \uu b \} \\
 &= \bigcap_{b \in B} \{ (d,e) \in D \times D \mid d \not\in \uu b \lor e \in \uu b \} \\
 &= \bigcap_{b \in B} ((D \setminus \uu b) \times D \cup D \times \uu b).
\end{align*}
As $B$ is countable and $\uu b$ is Scott-open, this is a set in $\Sigma_D \otimes \Sigma_D$. We therefore also have
\[
\{ (d,e) \in D \times D \mid d \geq e \} = \bigcap_{b \in B} (D \times (D \setminus \uu b) \cup \uu b \times D),
\]
by symmetry, which is clearly in $\Sigma_D \otimes \Sigma_D$, and the intersection of these sets, the diagonal $\{ (d,e) \in D \times D \mid d = e \}$, is therefore also, as are the complements of these sets. 
\end{proof}

\begin{lem}
\label{BorelTensorLemma}
Let $D$ be a dcpo with countable basis. Then $\Sigma_{D\times D} = \Sigma_D \otimes \Sigma_D$. 
\end{lem}
\begin{proof}
Let $B \subseteq D$ be a countable basis. Since the projection maps $\pi_i : D \times D \rightarrow D$ are Scott continuous, they are continuous with respect to the Scott topologies, and therefore measurable, so $\Sigma_D \otimes \Sigma_D \subseteq \Sigma_{D\times D}$. To show the opposite inclusion, first observe that $B \times B$ is a countable basis for $D \times D$ \cite[Proposition III-4.12 (i)]{gierz}, and therefore $(\uu (b_1 \times b_2))_{(b_1,b_2) \in B \times B}$ is a countable set of generators for $\Bo(D \times D)$ (Theorem \ref{ScottBorelPresentationTheorem}). As $\uu (b_1 \times b_2) = (\uu b_1) \times (\uu b_2)$, this shows that $\Sigma_{D \times D}$ is generated by elements of $\Sigma_D \otimes \Sigma_D$, and therefore $\Sigma_{D \times D} \subseteq \Sigma_D \otimes \Sigma_D$. 
\end{proof}

\end{document}